%% file: aanda.tex
\documentclass{aa}  
\usepackage{float}
\usepackage{graphicx}
\usepackage{txfonts}
\usepackage[colorlinks=true, linkcolor=blue]{hyperref}
\DeclareUnicodeCharacter{2212}{-} % otherwise returns a warning
\usepackage{adjustbox} % to adjust the size of a table
\usepackage{natbib}
\usepackage{url}

\begin{document} 

    \title{Characterising the properties of the atmospheric emission at Teide Observatory in the 10--20\,GHz range with QUIJOTE data}
    \titlerunning{Atmospheric emission at Teide with QUIJOTE}

    \author{Apolline Chappard \inst{1,2},
          José Alberto Rubiño-Martín \inst{2,3},
          \and
          Ricardo Tanausú Génova Santos\inst{2,3}\fnmsep 
          }
    \authorrunning{A. Chappard et al.}

   \institute{Institut d'Astrophysique Spatiale (IAS), CNRS, Bât 120 – 121 Univ. Paris-Saclay,
              ORSAY CEDEX, 91405, France\\
              \email{apolline.chappard@universite-paris-saclay.fr}
         \and
             Instituto de Astrofísica de Canarias (IAC), Organization, C/ Vía Láctea, La Laguna, E-38205, Tenerife, Spain 
        \and
        Departamento de Astrofísica, Universidad de La Laguna (ULL), La Laguna, E-38206, Tenerife, Spain
             }

   \date{Received XXX; accepted XXX}

% \abstract{}{}{}{}{} 
% 5 {} token are mandatory
 
  \abstract
   {QUIJOTE is a CMB experiment composed of two telescopes, QT1 and QT2, located at the Teide Observatory in Tenerife, Spain. The MFI instrument (2012–2018), installed on QT1, observed the sky at four frequency bands, namely 11, 13, 17, and 19\,GHz, with one degree angular resolution. Its successor, MFI2, began operations in early 2024 and operates in the same frequency bands.}
   {This paper has two main goals. First, we characterise the atmospheric conditions at the Teide Observatory to improve existing models at these CMB frequencies. Second, we carry out an empirical characterization of the  atmospheric turbulence using observations from both QUIJOTE MFI and MFI2. This work has implications for both atmospheric physics and CMB observations, and can be used for future reanalyses of MFI data, or in the preparation for upcoming instruments such as the Tenerife Microwave Spectrometer.}
   {We used data from GPS antennas, the STELLA observatory, and radio soundings to derive median profiles and distributions of key atmospheric parameters in the period 2012--2018. We then analysed MFI data to compute atmospheric structure functions at 17 and 19\,GHz. Using the full MFI database, we studied the correlation properties of the atmospheric signal by calculating the cross-correlation function of the time-ordered data between horns operating at the same frequency. Finally, we used MFI2 observations to study the atmospheric power spectrum, and compared it to the determination of the structure function based on MFI data.}
   {The water vapour density profile above the observatory can be well described by an exponential decay law, with a characteristic half-height of about 1000\,m. Median values of PWV in the 2012-2018 period are 3.3\,mm, with a 25th percentile of 2.1\,mm. 
   For high PWV conditions, we find that the structure function estimated with MFI data is consistent with the Kolmogorov turbulence model. The slope of the power spectrum of the atmospheric emission seems also consistent with the prediction of this model, although within a range of frequencies limited by the outer scale and by the instrument noise. Furthermore, through the study of the coherence length in the correlation function, we confirmed that atmospheric conditions remain stable for a period of about 1-2 hours. These results show that the Teide Observatory has an atmospheric behaviour comparable to that of the ACT site, although with higher integrated precipitable water vapour due to its lower altitude.}
   {}


   \maketitle

% original: Our results show that the Teide Observatory has similar atmospheric behaviour to that of the ACT site, though with higher humidity due to its lower altitude. For high PWV conditions, we confirm the Kolmogorov turbulence model through structure function analysis. Regarding the power spectrum, we find evidence of the Kolmogorov spectrum and the turbulence outer scales. Additionally, we confirmed that atmospheric conditions remain stable over a time period of about 1-2 hours.
%-------------------------------------------------------------------
\section{Introduction}

% CONTEXT
Observations of the Cosmic Microwave Background (CMB) offer a unique window into the fundamental physics of the early Universe \citep{Bennett2013,2020A&A...641A...6P}. 
One of the primary goals of upcoming CMB research is the detection of a possible primordial gravitational wave background component generated during the inflationary epoch \citep{guth1981inflationary,  kofman1994reheating}. 
%primordial B-modes as predicted by the theory of inflation, in which the size of the universe expanded exponentially by 50 to 60 e-fold in a very brief time period of about 10$^{-36}$ seconds after the Big Bang  \citep{guth1981inflationary, kofman1994reheating}. 
%This extremely energetic process might have created a background of gravitation waves that 
These signals could be indirectly observed today through their imprint on the CMB polarisation maps as B-modes, with an amplitude parameterised by the tensor-to-scalar ratio $r$, which is related to the energy scale of inflation \citep{Zaldarriaga1997,Kamionkowski1997}. 
The current best upper limit on the tensor-to-scalar ratio is $r < 0.032$ at 95 per cent confidence level, based on the combined analysis of BICEP/Keck and Planck data \citep{Tristram2022}. 
Upcoming ground-based experiments such as the Simons Observatory \citep{SO2019}, along with space missions like LiteBIRD \citep{litebird2023probing}, are expected to significantly improve these constraints in the coming years, thanks to remarkably better instrument sensitivities, control of systematics effects and correction of Galactic foreground emission. 
%
%To detect smaller B-mode signals, we need to increase the sensitivity of the instrument and improve the characterisation of the CMB foregrounds, such as the synchrotron emission and the thermal radiation from the interstellar dust grains \citep{Bennett2013, Planck2020d}. 

For ground-based CMB observatories, one of the main challenges in terms of systematic effects is atmospheric emission.
%, which dominates over the CMB anisotropies. 
%
Although this emission is expected to be largely unpolarized \citep[e.g.,][]{2003NewAR..47.1159H, Battistelli2012, 2019ApJ...870..102T, 2020ApJ...889..120P}, the most harmful effect may come from instrumental effects that lead to intensity-to-polarisation leakage, as bandpass mismatch or non-ideal half-wave plates.

\begin{figure}[t!]
    \centering
    \includegraphics[width=9cm]{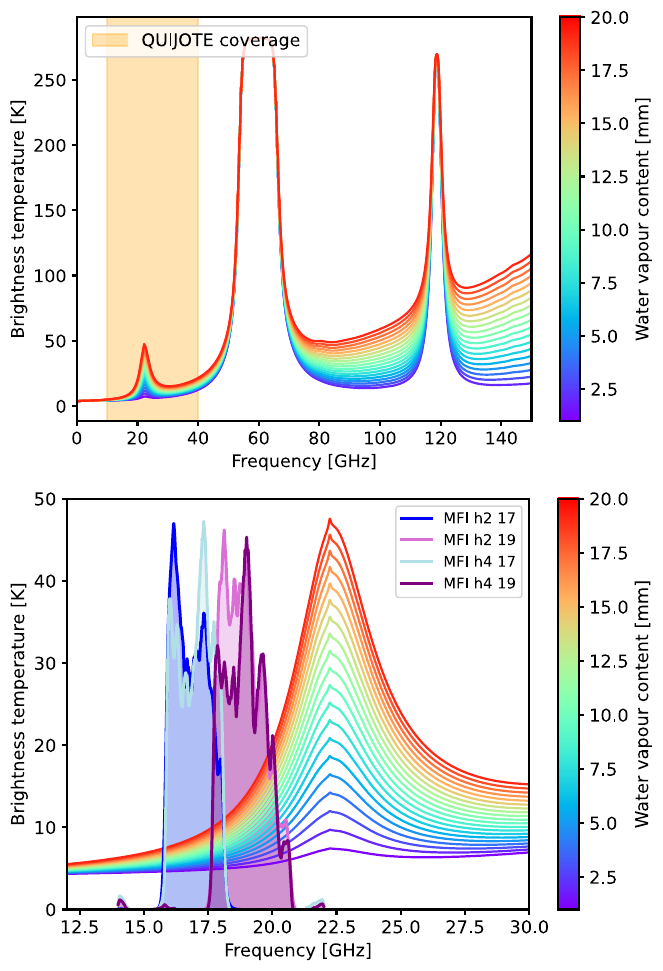}
    \caption{Top: brightness temperature of the atmosphere as a function of frequency between 1 and 150\,GHz for different water vapour content computed with the software \textit{am} \citep{Paine}, using the atmospheric conditions at the Teide Observatory. The frequency domain explored by the various QUIJOTE instruments is shown in orange.  Bottom: same figure zoomed in between 15 and 25\,GHz. The bandpasses of the QUIJOTE MFI horn 2 and 4 are shown in arbitrary units in dark blue and purple for horn 2 at 17\,GHz and 19\,GHz, respectively, and in cyan and pink for horn 4 at 17\,GHz and 19\,GHz, respectively.} 
%    Those bandpasses are located on the left of the water vapour emission line at 22\,GHz to avoid atmospheric contamination as much as possible.}
    \label{fig:TB_atmo}
\end{figure}

In the CMB relevant frequencies for ground-based experiments (say 10--200\,GHz), atmospheric contamination is mainly caused by two emission lines of water vapour at 22\,GHz and 183\,GHz, and two lines of molecular oxygen at 60\,GHz and 120\,GHz \citep[e.g.][]{ATM, Paine}. This is illustrated in the top panel of Figure~\ref{fig:TB_atmo}, which shows the brightness temperature of the atmosphere as a function of frequency for different water vapour content simulated with the \textit{am} software \citep{Paine}. 
%
%As the water vapour content in the atmosphere changes,  the amount of radiation that the atmosphere absorbs and re-emits changes as well. This leads to variations in the observed brightness temperature, especially at the frequencies of the emission lines. Hence, even small changes in the distribution of water vapour can introduce significant fluctuations in the observed atmospheric signal. Moreover, 
For observations of CMB anisotropies, water vapour is the most problematic source of contamination, as it has a highly variable and inhomogeneous concentration in time and space due to the turbulent behaviour of the atmosphere \citep{Tatarski1961}. 
This introduces significant variations in the atmospheric signal, with characteristic amplitudes and spatial scales that depend on the physical conditions of the observing site (air density, water vapour content, temperature, pressure, etc.). In the time-ordered data, this signal effectively appears as correlated noise, which is difficult to distinguish from the true CMB signal. Additionally, wind can shift atmospheric structures, introducing non-stationary effects.
All these effects project onto large angular scales in the sky maps reconstructed by the instruments \citep[e.g.,][]{Morris2022}, thereby limiting our ability to characterise the so-called reionisation bump, whose detection is extremely challenging from the ground.

% STATUS OF THEORY NOW
%The current method for subtracting these fluctuations from the CMB involves identifying their spatial correlations across different instrument channels, whether observing at the same or different frequencies. These fluctuations exhibit intensity variations that follow a characteristic scaling factor across frequencies. [RW]
%Although the atmosphere is generally expected to be unpolarised, some measurements suggest the presence of polarisation, possibly because of ice crystals. More importantly, the brightness of the atmosphere causes significant temperature-to-polarisation leakage, which can severely affect B-mode measurements. 

Modelling those atmospheric effects is inherently complex. A reference theoretical framework with a 3D model of the atmosphere was established by \citet{Church1995}, an approach being used and updated by current data simulation frameworks. 
Further studies were using the inputs from specific CMB experiments conducted at the Atacama Desert site (Chile) and the South Pole. At the Atacama desert site, \citet{Errard2015} used POLARBEAR data \citep{polarbear} to parameterise a 3D model of turbulence, generalising the approach by \cite{Church1995}. More recently,  \citet{Morris2022, Morris2024} used Atacama Cosmology Telescope (ACT) data \citep{Thornton2016} to establish a model of the atmosphere as a discrete set of emissive turbulent layers and to present statistics on the fluctuations of atmospheric emission. 
At the South Pole site, \citet{Bussmann2005} used data from the Arcminute Cosmology Bolometer Array Receiver (ACBAR) \citep{Kuo2004} to measure the brightness fluctuations produced by the atmosphere, and \citet{Coerver2025} presented the fluctuations in linearly polarised emission from the atmosphere based on the use of data from SPT-3G. 

%OT 
This paper focuses on the observational characterisation of the atmospheric properties at the Teide Observatory (OT), located at latitude $28^{\circ}18^\prime04^{\prime\prime}$ North and longitude $16^{\circ}30^\prime 38^{\prime\prime}$ West and at an altitude of 2,400\,m in Tenerife, Canary Islands. OT is a site with excellent observing conditions, well characterised over the years by a dedicated Sky Quality Team. 
The median Precipitable Water Vapour (PWV), defined as the total water vapour in a column of atmosphere integrated from the surface to the top with a unit cross-sectional area, is 3.5\,mm \citep{mfiwidesurvey, Castro2016}. In addition to its relatively low PWV, OT also benefits from a highly laminar atmospheric flow, which further improves observing conditions. This peculiarity arises from two main factors: the presence of a temperature inversion layer at an altitude of around 1500\,m and persistent trade winds from the North. The water vapour contained in these humid winds undergoes condensation at this level, creating a cloud layer with a sharp end, and resulting in a very dry and stable atmosphere above.
%The inversion layer is a well-known feature at the latitude of Tenerife, which lies at the intersection of the Hadley and Ferrel atmospheric cells. This causes a downward airflow from higher altitudes, leading to atmospheric compression and the formation of the inversion layer. 
%As a result, despite not having the lowest PWV among candidate sites, OT remains highly competitive for CMB observations due to its stable and dry atmospheric conditions.

%In this paper, we are using data from QUIJOTE, which operates at 10-40\,GHz where the atmosphere isn't too bright. This study will allow to explore how good this observatory could be to perform CMB observations at higher frequencies (and you should mention here also GroundBIRD and Strip). And, in addition, your study will also provide important information for other CMB experiments in general, like SO, CLASS, and others...

OT  has a long tradition in CMB research since 1984.
%due to its dry and stable atmosphere. 
Multiple CMB experiments have been conducted at this observatory, including the Tenerife experiment \citep{gutierrez2000tenerife}, %JBO-IAC interferometer \citep{dicker1999cosmic}, 
the JBO-IAC 33\,GHz interferometer \citep{melhuish199933}, the Very Small Array (VSA) \citep{watson2003first}, or COSMOSOMAS \citep{fernandez2006observations}. 
For this paper, we use data from the Q-U-I JOint TEnerife (QUIJOTE) experiment \citep{mfiwidesurvey}, which is operating from the OT since 2012. QUIJOTE is a scientific collaboration between the Instituto de Astrofísica de Canarias (IAC), the Instituto de Física de Cantabria (IFCA), the Universities of Cantabria, Manchester and Cambridge, and the IDOM company. It consists of 3 instruments covering the range of $10$--$40$\,GHz mounted on two Cross-Dragone $2.25$\,m primary aperture telescopes.
%; see \citet{QT1} and \citet{QT2b} for a complete description of the instruments. The primary scientific goals of the QUIJOTE experiment are 1) to detect the CMB B-modes if they have an amplitude greater than or equal to $r=0.05$ and 2) to provide information on the astrophysical microwave foregrounds. 
The analyses performed in this paper are based on data collected by two QUIJOTE instruments mounted on the first QUIJOTE telescope, both covering the 10--20\,GHz frequency range: the Multi-Frequency Instrument (MFI) \citep{MFIstatus12} that was operative between November 2012 and October 2018, and the Second Multi Frequency Instrument (MFI2) \citep{MFI2}, that started operating at the beginning of 2024. 
In particular, we make extensive use of the 17 and 19\,GHz detectors of MFI and MFI2 (see Fig.~\ref{fig:TB_atmo}), due to their proximity to the 22\,GHz water line. Both instruments have two independent horns/receivers observing those bands, and thus, cross-correlation analyses will also be used for our study. 

%The atmospheric emission at QUIJOTE frequencies is relatively low.
% WHAT WE DO
This study aims to improve our understanding of the atmospheric signal at OT, providing detailed information for future analyses at these frequencies or higher.
To this end, we first obtain the average atmospheric conditions at OT using data from sounding stations (water vapour density, temperature, pressure) and GPS measurements (PWV). As a reference period for this study, we use 2012-2018, which is the time span when the QUIJOTE MFI instrument was operative and carried out the MFI wide survey \citep{mfiwidesurvey}. 
Then, we used the QUIJOTE data at 10--20\,GHz to characterise the scale-dependent properties of the atmospheric emission at these frequencies. QUIJOTE MFI and MFI2 data are used to measure the spatial atmospheric structure function and the power spectrum, to verify the theoretical prediction under the Kolmogorov theory of turbulence \citep{kolmogorov1941}. The latter is also used to obtain an indication of the size of the atmospheric turbulence outer scale $L_0$. 
%We compared the results with the prediction of the Kolmogorov theory. Additionally, 
Finally, we computed the cross-correlation function of the atmospheric signal to estimate the atmospheric coherence length and to verify the validity of the assumption used in \citet{mfiwidesurvey} for atmospheric removal in the intensity maps. That analysis assumed that the atmosphere remained stable over a one-hour period to reconstruct large-angular-scale atmospheric patterns along the azimuth direction.

%I think it is better to insert this paragraph after your description of QUIJOTE. It is better to first describe QUIJOTE, which is a CMB experiment, and then say that the analysis tools, techniques, ant theoretical formalism that you have used in this study are the same as have been used in previous papers using data from CMB experiments.

The paper is organised as follows: Section~\ref{section2} presents the data sources used in this analysis. Section~\ref{section3} describes the average atmospheric conditions at the Teide Observatory. Section~\ref{section4} gives an explanation of the basic aspects of the processing of QUIJOTE data. The different analyses and results are presented and discussed in sections~\ref{section:CL}, \ref{section6} and \ref{section:MFI2}, while the conclusions of this work are presented in ~\ref{section8}.

%Then, section \ref{section4} gives a brief overview of the theoretical aspect of atmospheric behaviour. Section \ref{section5} presents the analysis and the results. The main conclusions of this work are given in Section \ref{section6}

%--------------------------------------------------------------------
\section{Input Data} 
\label{section2}

Our analyses are based on multiple data sources, including sounding stations, GPS antennas, weather stations and microwave data from the QUIJOTE MFI and MFI2 instruments.

\subsection{Sounding Stations, Weather Sensors, and Auxiliary Equipment at the Teide Observatory}
\label{sec:stations}

To determine the vertical profiles of temperature, pressure, and water vapour density above the OT, we use data from radio soundings launched from the nearby Güímar sounding station (World Meteorological Station \#\,60018), located at latitude 28°19'06.0" North and longitude $16^{\circ}22^{\prime}56.0^{\prime\prime}$ West, approximately 15\,km east of the Observatory, at an altitude of 115\,m above sea level. These soundings are launched twice per day (at noon and at midnight) and provide measurements of the atmospheric pressure $P$, temperature $T$, relative humidity $\varphi$, and mixing ratio $\zeta$ as a function of height, at height intervals of $\sim 100$\,m and up to about 30,000\,m. Data are publicly available at the University of Wyoming Atmospheric Science Radiosonde Archive\footnote{\url{https://weather.uwyo.edu/upperair/sounding.shtml}}.

Effective monitoring of the PWV at the observatory site is essential for atmospheric correction. GPS antennas provide a reliable method for estimating the integrated PWV along the atmosphere. The technique is based on studying the delays induced in the GPS signal due to atmospheric refraction \citep{bevis1994, Castro2016}. 
To characterise the distribution of PWV at OT, we use data from the EUREF GPS station (IZAN), located at the Centro de Investigaci\'on Atmosf\'erica de Izaña (CIAI, Agencia Estatal de Meteorología; AEMET\footnote{\url{https://izana.aemet.es/?lang=en}}), 1400\,m away from QUIJOTE at an altitude of 2367\,m, which is 13\,m below the altitude of QUIJOTE. This altitude offset leads to a difference of only 0.024\,mm in PWV as derived from radio sounding. The GPS station produces one PWV measurement every 15 min (rapid orbits) or every 60 min (precise orbits), and we have these data available throughout all QUIJOTE MFI wide-survey observations (2012 to 2018).

Finally, statistics of other meteorological parameters, such as wind speed and direction, are analysed using data from the weather station attached to the STELLA Robotic Observatory \citep{stella2004}, operated by the Leibniz Institute for Astrophysics Potsdam\footnote{\url{https://stella.aip.de/}}, and located approximately 100\,m from the QUIJOTE telescopes. While other meteorological stations are available at the site\footnote{\url{https://www.iac.es/en/observatorios-de-canarias/teide-observatory/weather}}, we selected this one due to its proximity to the QUIJOTE telescopes. The corresponding public data can be retrieved online\footnote{\url{http://stella-archive.aip.de/stella/status/getdetail.php?typ=3}}.

\subsection{QUIJOTE MFI data} \label{section:filtering}

%\begin{table}
%    \centering
%    \begin{tabular}{c|cccc}
%         & Horn 1& Horn 2 & Horn 3 & Horn 4    \%\ \hline
%         Low frequency & 11.1 & 16.8 & 11.1 & 16.8 \\
%         High frequency & 12.9 & 18.8 &  12.9 & 18.8  \\
%         \hline
%    \end{tabular}
%\caption{Effective QUIJOTE MFI central frequencies $\nu_0$ in GHz for the 4 horns of the instrument. Each horn measured the sky in a low-frequency band and a high-frequency sub-band.}
%    \label{tab:freq_band}
%\end{table}

MFI \citep{MFIstatus12} was the first instrument installed on the first QUIJOTE  telescope (QT1). It measured the intensity and polarisation of the sky at frequencies 11, 13, 17, and 19\,GHz from November 2012 to October 2018. It consisted of four horns with their corresponding polarimeters, two of them observing the sky at 11--13\,GHz and the other two at 17--19\,GHz. All frequency bands had an approximate bandwidth of 2\,GHz. The analyses presented in this paper are based on horns 2 and 4 that, covering the 17--19\,GHz band, are the most sensitive to the water vapour emission thanks to their proximity to the 22\,GHz line. Taking advantage of having two different horns observing at the same frequencies (effective frequencies of 16.8 and 18.8\,GHz) and with different $1/f$ noise properties, we perform cross-correlation analyses that allow separating the atmospheric $1/f$ noise from that introduced by the instrument. 
%These two horns were also chosen because they are the most sensitive to atmospheric emission. The effective central frequencies for those two high frequency sub-bands are 16.8 and 18.8\,GHz, both for horn 2 and 4 \citep{mfiwidesurvey}.

MFI operated for approximately 6 years, with an observing efficiency of almost 50 per cent, resulting in 26,000 hours of raw data.
Approximately one third of that observing time was devoted to the MFI wide survey \citep{mfiwidesurvey}. Here we use that dataset.  
Although individual observing sessions typically lasted 24 hours, for practical reasons related to file management, one file containing calibrated time-ordered data (CTOD) is generated every 8 hours (so one day of observations typically has 3 CTOD files). In total, there are 1233 such data files contributing to the published MFI wide survey maps. Hereafter, we will refer to each of these 8-hour data files as an “observation”. 

The wide survey observations were carried out in the so-called nominal mode configuration, where the telescope scanned the sky in azimuth circles at a fixed elevation. Different elevations were used during the wide survey in order to optimise the uniformity of integration time per pixel. These elevations were $30^\circ$, $35^\circ$, $40^\circ$, $50^\circ$, $60^\circ$, $65^\circ$ and $70^\circ$. The scan speed was $v_{az} = 6$\,deg/s before January 9th 2014. This speed was increased to $v_{az} = 12$\,deg/s after that date in order to improve the mitigation of $1/f$ noise.  We note that the acquisition dates of the data file are irregular, as there are extended periods with no observations in nominal mode \citep[see ][for more details]{mfiwidesurvey}. 

Each MFI frequency band produced four outputs, called channels. Since each horn had two frequency bands, each horn produced eight channels in total. The channels are denoted $x$, $y$, $x + y$, and $x − y$, and they measure different combinations of the Stokes parameters $I$, $Q$ and $U$. For this study, we focus on the intensity signal only, obtained from the sum of the correlated channels $x + y$ and $x − y$.

%I am not sure this figure should be introduced. Maybe remove it. A table with some basic statistics of how much data have been flagged might be enough.

%[TO BE MOVED BELOW]
%In the second part of this paper (see Section \ref{section:CL}), we analyzed the cross-correlation function of the signal of horn 2 and horn 4 at 17\,GHz and 19\,GHz using MFI wide survey observations to analyse the correlation of the atmospheric signal across different horns. For this analysis, the number of flagged samples (the flagging of QUIJOTE MFI is described in \citet{mfiwidesurvey}) in each observation was crucial because it significantly affects the correlation function. Flagged samples were set to zero, which reduces the overall correlation by artificially lowering the product of data pairs. As a result, a higher proportion of zeros biases the correlation function downward, making it appear weaker than it truly is. Therefore, it was essential to exclude from the analysis the observations containing too many flagged samples.  After analysing the impact of the flagging on the cross-correlation function, we decided to keep files containing less than 30\% of flagged samples to compute the cross-correlation function, as the effects below this value were negligible. We therefore removed 711 observations from the 1233 MFI observations on this flagging condition for this analysis, which is about 58\% of the observations. More details about this data selection are given in the Appendix \ref{section:data_sel}

%--------------------------------------------------------------------
\subsection{QUIJOTE MFI2 data} \label{section:QUIJOTE_MFI2}

MFI2 \citep{MFI2} is an upgraded version of the original MFI instrument, using state-of-the-art detectors, a simplified radiometer architecture, and a digital back-end based on FPGAs for RFI mitigation.
The instrument began taking data in early March 2024, initially using the former MFI back-end as a provisional solution. It consists of five polarimeters: three operating in the 10--15\,GHz sub-band and two in the 15--20\,GHz sub-band. Commissioning observations of MFI2 confirm that the new instrument is between two and three times more sensitive than the original MFI, as expected.

%In this paper, we analysed data collected by MFI2 in dedicated observations for atmospheric studies. Those were obtained when the telescope was kept at a fixed azimuth and elevation to compute the atmospheric cross-power spectral density (see Section \ref{section:MFI2}). 
%Keeping the telescope fixed allows a clean measurement of the atmospheric power spectrum. This is because rotating the telescope introduces a sinusoidal pattern in the time output signal, as shown in Figure \ref{fig:signals_comp}. This pattern is caused by spatial variations in atmospheric brightness, primarily due to uneven distributions of water vapour. For instance, if one direction in the sky has more water vapour, it will appear brighter. As the telescope rotates in a circular scanning pattern, it repeatedly passes through that brighter region. This causes the output signal to rise and fall periodically, peaking when the telescope points toward the bright region and dropping when it points away. The result is a sinusoidal oscillation in the signal, following the telescope's rotation, which then appears in the power spectrum as peaks of power at the telescope's rotation frequency and its harmonics. By keeping the telescope fixed, we avoid this scanning-induced oscillation and obtain a stable time output of the atmospheric fluctuations for the computation of the power spectrum.

In this paper, we analyse data from a set of dedicated MFI2 observations carried out in July 2024 to characterise atmospheric emissions. A total of 19 scans were performed, each lasting approximately 20 minutes, with the telescope pointing at a fixed local coordinates (see Table~\ref{tab:list_MFI2} of the Appendix \ref{ap_MFI1} for details). Three different elevations were used: $30^\circ$, $60^\circ$, and $90^\circ$. These observations were specifically designed to enable computation of the atmospheric cross-power spectral density (see Sect.~\ref{section:MFI2}).
We note that the MFI wide survey observations cannot be used for this computation, as the spin-synchronous atmospheric patterns present in the data would appear superimposed on the actual atmospheric emission.

%For this study, we took 19 observations with MFI2 in July 2024 of about 20 minutes each at different telescope elevations, namely elevations 30°, 60°, and 90°. Table \ref{tab:list_MFI2} contains a summary of key parameters for each observation, including the telescope elevation, observation duration, date and time, as well as the atmospheric conditions at OT (PWV, wind speed, and wind direction) during the observations.

\section{Atmospheric conditions at Teide Observatory for 2012-2018} \label{section3}

 %We first present the median profiles of some atmospheric parameters using the data of the Güímar Sounding station \footnote{Radiosonde observations for World Meteorological Station \#60018 (Tenerife-Guímar, Canary Islands, Spain) (\url{https://weather.uwyo.edu/upperair/sounding.shtml})}. We show the distribution of PWV at OT using data from a GPS antenna \footnote{EUREF GPS station (IZAN), belonging to the Centro de Investigacíon Atmosférica de Izaña (CIAI, Agencia Estatal de Meteorología; AEMeT) (\url{https://izana.aemet.es/?lang=en})}. Finally, we present the distribution of wind speed and wind velocity at OT using data measured by the weather station attached to the STELLA telescope\footnote{STELLA is a robotic telescope, belonging to Leibniz-Institut für Astrophysik Potsdam, and installed at the Teide Observatory (\url{https://stella.aip.de/})}.

This section presents a statistical analysis of the atmospheric conditions at OT using the data sources described in Sect.~\ref{sec:stations}. 
As a reference period of time for the different studies, we consider the lifetime of the QUIJOTE MFI instrument, which operated from November 2012 to December 2018.

%%%%%%%%%%%%%%
\subsection{Atmospheric parameters from radiosonde data}

\begin{figure*}
    \centering
    \includegraphics[width=18cm]{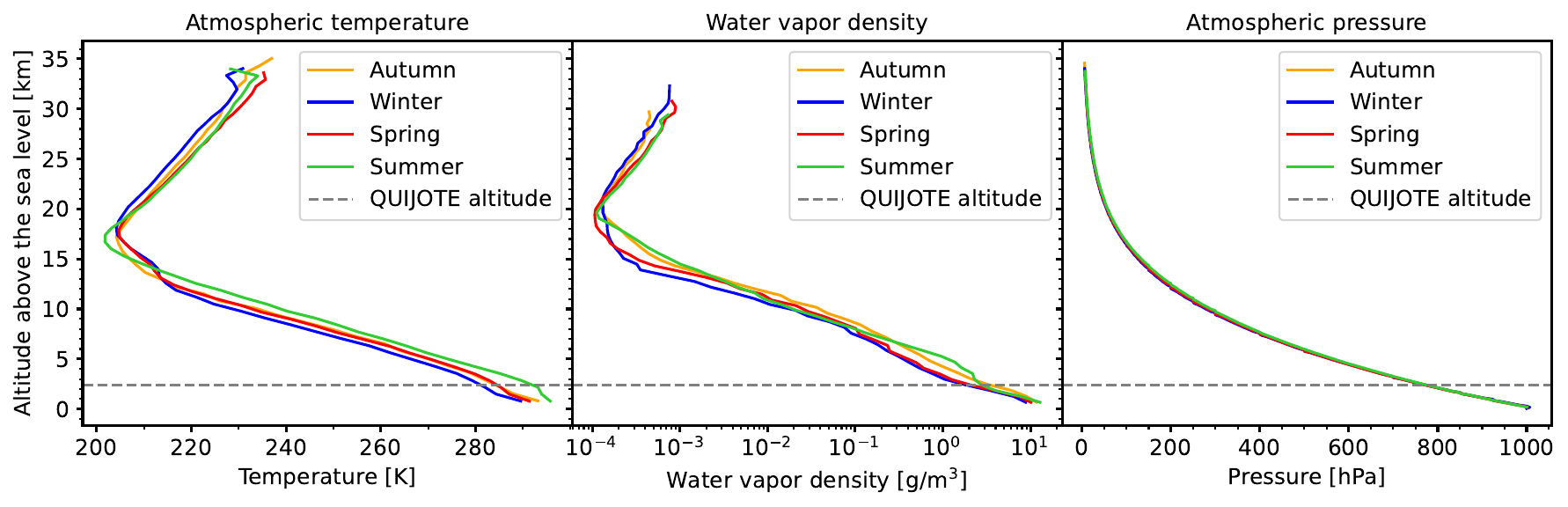}    
    \caption{Median seasonal profiles of atmospheric temperature (left), water vapour density (middle), and atmospheric pressure (right) measured by radio-sounding, spanning altitudes from 105\,m to 33\,km. In the three panels, the median profiles for autumn (orange), winter (deep blue), spring (red) and summer (green) are shown. The altitude of QUIJOTE is shown as a grey line.}
    \label{fig:full_profiles}% gives a warning because I call the reference before defining it 
\end{figure*}

\begin{figure*}
    \centering
    \includegraphics[width=18cm]{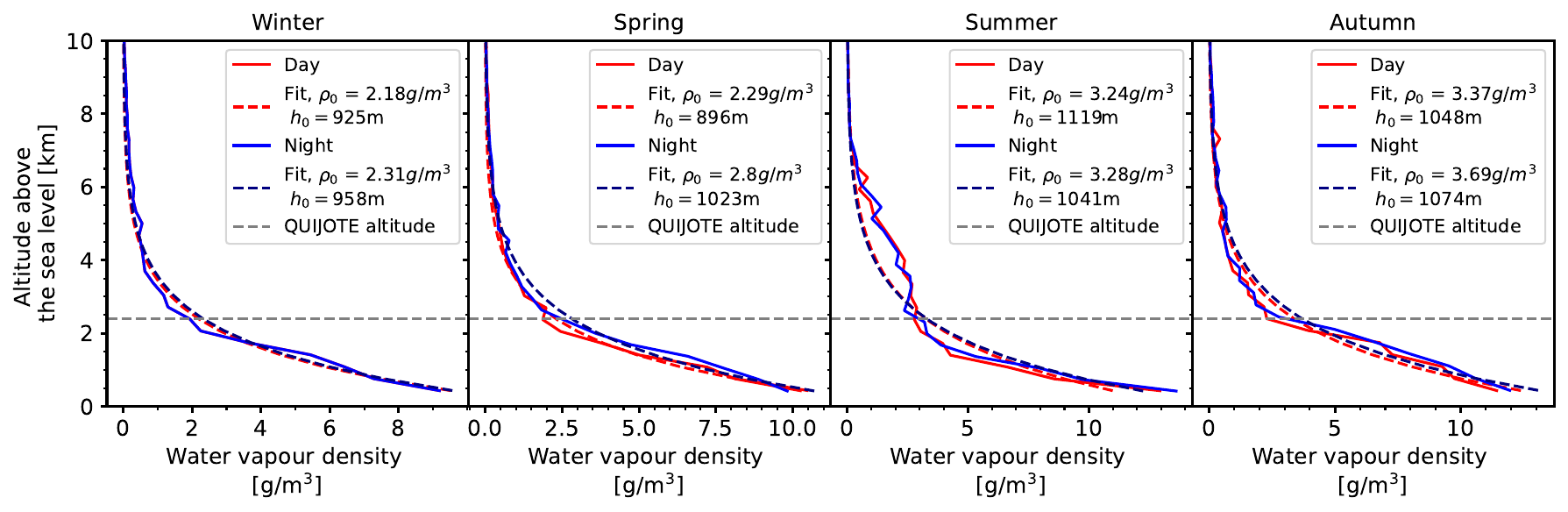}
    \caption{Binned median water vapour density as a function of the altitude above Güímar for the four seasons during the day (red curves) and night (blue curves) calculated using equation \ref{eq:rho}. The QUIJOTE altitude is denoted as a grey line. The full lines represent the binned data, and the dashed lines represent the exponential fits obtained with equation \ref{eq:exp_decay}.}
    \label{fig:med_rho}
\end{figure*}

%  We used equation \ref{eq:rho} to calculate the water vapour densities from the radio-sounding datasets measured between January 2012 and December 2018. From the fits, we can extract the half-heights of each profile, which are around a kilometre, and the water vapour densities at the telescope altitude, which are about 2.3\,g/m$^3$ in winter and spring and 3.3\,g/m$^3$ in summer and autumn. Moreover, the water vapour density is generally slightly lower during the day.

We used 7 years of radio sounding data (see Sect.~\ref{sec:stations}) measured between January 2012 and December 2018 (covering the entire time period of the MFI wide survey) to calculate the median water vapour density profiles for all four seasons during the day and the night. From the measured variables described in Sect.~\ref{sec:stations}, we can infer the water vapour pressure $P_\text{H$_2$O}$, which is the pressure exerted by molecules of water vapour in gaseous form, using equation 3.1 and the coefficients from \cite{flatau1992}. From there, we derive the water vapour density as
\begin{equation} 
\label{eq:rho}
\rho_\text{H$_2$O} = P_\text{H$_2$O} \frac{M}{R T}, 
\end{equation}
\noindent
with $M = 0.01801528$\,kg/mol the water molar mass, and $R = 8.314472$\,m$^3 $Pa/(K mol) the gas constant.

The atmospheric data recorded by radiosondes are not uniformly distributed in altitude, as the measurement heights vary slightly between soundings, leading to irregular vertical sampling. To construct the median vertical profiles, we applied a binning procedure: the data were grouped into fixed-altitude bins, and the median value was computed within each bin.
%This approach allows for a smoothed and statistically meaningful representation of the profiles. 
Figure~\ref{fig:full_profiles} shows the median atmospheric temperature, atmospheric water vapour density and atmospheric pressure profiles for the four seasons. We use here the full radio-sounding altitude database, extending up to 30\,km, for a direct comparison with the results of \cite{otarola2018}. Seasons are defined as follows: “summer” covers 21 June to 21 September, “autumn” from 22 September to 21 December, “winter” from 22 December to 19 March, and “spring” from 20 March to 20 June. 

The temperature profiles depicted in the left panel of Fig.~\ref{fig:full_profiles} show an almost linear decrease with altitude up to the tropopause. 
%, which is the boundary between the troposphere (the lowest atmospheric layer) and the stratosphere. 
At this point, an inversion layer occurs: instead of continuing to decrease, the temperature begins to increase with altitude. This happens in the stratosphere due to the absorption of the Sun's ultraviolet (UV) radiation by the ozone layer. The water vapour density profile (central panel in Fig.~\ref{fig:full_profiles}) shows an exponential decrease with altitude below the tropopause. This rapid decrease happens because most water vapour originates from the Earth's surface (through evaporation) and condenses as it rises due to decreasing temperature and pressure. However, in the stratosphere, there is a slight increase in water vapour density due to processes such as the photochemical oxidation of methane (CH4), which produces additional water vapour at high altitudes. Moreover, we can see more water vapour near the telescope altitude (grey vertical line) in summer compared to other seasons. The atmospheric pressure profile (right panel in Fig.~\ref{fig:full_profiles}) follows an exponential decay consistent between all seasons. Indeed, unlike temperature and water vapour, which are influenced by seasonal changes, pressure at a given altitude is primarily controlled by gravity and the total mass of the atmosphere, which remains nearly constant throughout the year. 

Figure~\ref{fig:med_rho} shows the water vapour density profiles, but extending up to 10\,km, where the main contribution to the integrated water vapour density occurs. In this range, the observed  profiles follow approximately an exponential decay law, which can be fitted as 
\begin{equation}
\label{eq:exp_decay}
\rho(h) = \rho_0 \exp \left( -\log(2) \cdot \frac{h - 2400\text{\,m}}{h_0} \right),
\end{equation}
with $\rho_0$ being the water vapour density at the QUIJOTE telescope altitude (i.e., 2400\,m) and $h_0$ the half height, which is the height at which the density falls to half its initial value $\rho_0$. We fitted this formula to all eight profiles (four seasonal profiles during the day and four during the night). Here, “day” refers to the measurements taken at noon UTC+0, and “night” to those taken at 00:00 UTC+0. 

Our best-fit values are indicated in the labels of Fig.~\ref{fig:med_rho}. The typical half-height is about 1\,km for all seasons at night and day. This indicates that 50 per cent of the water vapour is concentrated within the first kilometre of the atmosphere, with most signal contamination occurring within this nearby region. This value is comparable to that obtained with ACT data for the Chajnantor site \citep{Morris2022}.
The values for $h_0$ are slightly higher during the night, except for summer. However, the exponential function provides a poor fit for the observed profiles. It is important to note that these profiles represent the median water vapour density based on a 7-year data set. Daily variations in the profile caused by atmospheric turbulence and different weather conditions are significant. This can be confirmed through the calculation of the 1 $\sigma$ region of the profile (not shown in this paper for clarity and due to space limitations), which is quite high, and indicates significant day-to-day variability in water vapour conditions. The typical water vapour densities at the altitude of the QUIJOTE telescope $\rho_0$ during the days are about 2.2\,g/m$^3$ in winter, 2.3\,g/m$^3$ in spring, 3.2\,g/m$^3$ in summer and 3.4\,g/m$^3$ in autumn. During the night, the values are approximately 2.3\,g/m$^3$ in winter, 2.8\,g/m$^3$ in spring, 3.3\,g/m$^3$ in summer and 3.7\,g/m$^3$ in autumn. The values are, in general, lower during the day for all four seasons.

%--------------------------------------------------------------------
\subsection{PWV at the Teide Observatory from GPS station} \label{section:Izana}

\begin{figure}
\centering
\includegraphics[width = 9cm]{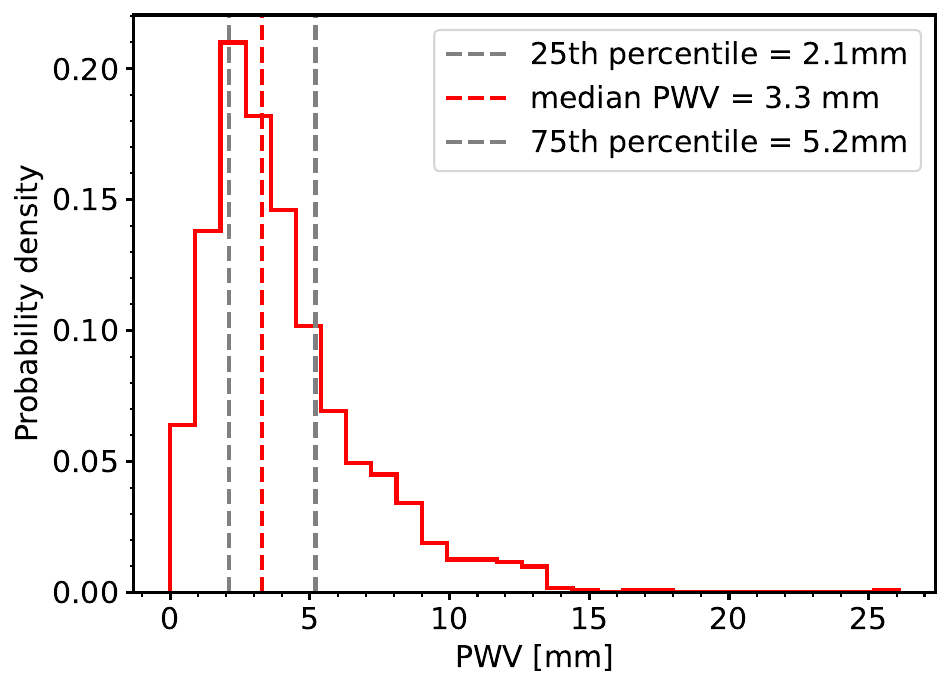}
\caption{Distribution of PWV measurements recorded during each QUIJOTE-MFI wide survey observation (period 2012-2018). Each PWV value corresponds to the median PWV during a single observation. The overall median PWV for the entire wide survey is approximately $3.3$\,mm. The first quartile indicates that 25\,\% of the observations have PWV values below 2.1\,mm. Similarly, the third quartile indicates that 75\,\% of the observations have PWV values below 5.2\,mm during the QUIJOTE MFI survey.}
    \label{fig:hist_PWV}
\end{figure}

The PWV is defined as the integral of the water vapour density $\rho_\text{H$_2$O}$ over a column of the atmosphere of height $z_\text{max}$ and unit cross-section and gives the total quantity of water vapour in that column. 
%It is expressed as:
%\begin{equation}
%PWV = \int_{z_0}^{z_\text{max}} \rho_\text{H$_2$O}(z) \text{d}z.
%\end{equation}
%
%\noindent
%Effective monitoring of the PWV at the observatory site is essential for atmospheric correction. GPS antennas provide a reliable method for estimating the integrated PWV along the atmosphere. The technique is based on studying the delays induced in the GPS signal due to atmospheric refraction \citep{bevis1994, Castro2016}. On QUIJOTE site, the PWV is measured by the EUREF GPS station (IZAN), belonging to the Centro de Investigacíon Atmosférica de Izaña (CIAI, Agencia Estatal de Meteorología; AEMeT), located 1400\,m away from QUIJOTE at an altitude of 2367\,m, 13\,m below the altitude of QUIJOTE. This altitude offset leads to a difference of only 0.024\ mm in PWV as derived from radio sounding. The GPS station produces one PWV measurement every 15 min (rapid orbits) or every 60 min (precise orbits), and we have these data available throughout all QUIJOTE wide-survey observations. 
For each observation in the MFI wide survey (on average 8\,h long), we calculated the median PWV during the observation using data from the Rapid Orbit observations measured by the IZAN station (see Sect.~\ref{sec:stations}). Thus, we end up with one representative PWV value per observation. Figure~\ref{fig:hist_PWV} shows the distribution of these PWV values, with an overall median of 3.3\,mm across the entire survey. 
These values are consistent with those found by \cite{Castro2016} for the OT.  
For comparison with other observatories, the PWV was around 1.1\,mm in winter and 2.3\,mm in summer at the ACT site \citep{Thornton2016} in the Atacama Desert, Chile, as measured by ERAS between January 1980 and January 2021 \citep{Morris2022}.
% site Ricardo report: Study of the Teide Observatory sky opacity using QUIJOTE data, and comparison with PWV measurements using GPS and other techniques 

%--------------------------------------------------------------------
\subsection{Wind parameters from STELLA station} \label{section:Stella}

\begin{figure}
\centering
\includegraphics[width = 9cm]{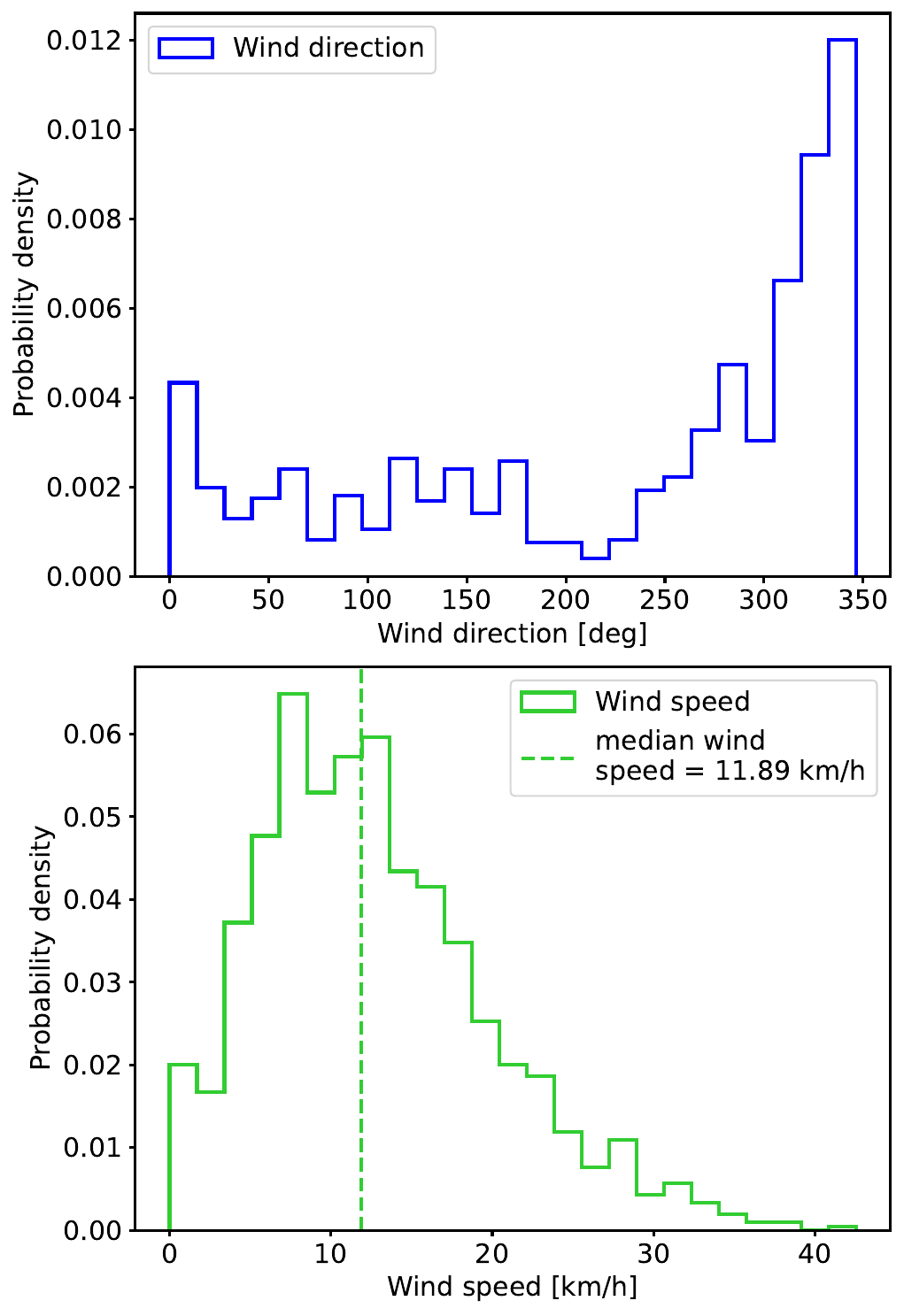}
\caption{Distribution of the wind direction (top) and the wind speed (bottom) recorded during each QUIJOTE-MFI wide survey observation (period 2012-2018). The wind velocity is measured in degrees as the direction the wind flows from north to east ($0^\circ$ is North, $90^\circ$ is East). The median wind speed was about 12\,km/h.}
\label{fig:hist_ws_wd}
\end{figure}

%This results in more rapid fluctuations in the signal, which can introduce additional correlated noise. 

Monitoring wind speed and direction at the observatory is crucial due to their influence on the atmospheric turbulence. Higher wind speeds will result in faster variations in atmospheric structure as the winds transport atmospheric turbulence more quickly across the telescope's field of view. Moreover, stronger winds inject more energy into the atmosphere, creating larger turbulence structures. The wind direction also impacts the atmospheric noise since it induces atmospheric signals that are directionally dependent. Understanding these impacts is crucial for accurate atmospheric signal characterisation. 

%At OT, the STELLA Robotic Observatory \citep{stella2004} provided us with the wind direction and wind speed\footnote{Public data available at \url{http://stella-archive.aip.de/stella/status/getdetail.php?typ=3}}. 
Figure~\ref{fig:hist_ws_wd} shows the distribution of wind speed and direction as measured by the STELLA meteorological station during the QUIJOTE MFI wide survey observations (period 2012 to 2018). For each individual MFI observation (approximately 8 hours long), the median of wind speed and direction was retrieved from the station's database, and the histogram was obtained for the 1233 observations.
Most of the time, the wind speed ranges between 8 and 15\,km/h with a median value of about 12\,km/h for the entire survey. We note that the QUIJOTE MFI observations are stopped and the telescope enclosure is closed when the wind speed exceeds 45\,km/h to prevent damage to the instruments. Thus, there are no data points beyond that value. At the OT, the wind direction is predominantly between $300^\circ$ and $360^\circ$ (north to north-west). Tenerife’s permanent trade winds, originating from the Azores high-pressure system, blow from the north-east and are slightly distorted at the QUIJOTE site due to the local orography.

%%%%%%%%%%%%%%%%%%%%%%%%%%%%%%%%%
% Preparation of MFI data
\section{Preprocessing of the QUIJOTE MFI data}
\label{section4}

% MFI data
Some of the analyses carried out in this paper with QUIJOTE MFI wide survey data require a pre-processing of the calibrated timelines (CTOD files) in order to guarantee that the remaining signal is dominated by atmospheric emission.
Here, we describe the expected signal contributions to the timelines and the procedures for preprocessing those CTOD timelines. 

\subsection{Signals in QUIJOTE MFI data and CTOD preparation} 

The intensity signal measured by the QUIJOTE MFI is a superposition of different components: 1) the astrophysical sky signal; 2) the CMB dipole that was removed from the public QUIJOTE maps; 3) the atmospheric signal; 4) possible radio frequency interference (RFI) contamination; and 5) instrumental noise. Since our analysis is focused on the atmospheric signal, we carry out a preprocessing of the CTOD files directed to remove the astrophysical sky, the CMB dipole and the RFI. 

The astrophysical sky signal is removed by re-projecting the final QUIJOTE maps  \citep{mfiwidesurvey} in the time domain. 
A CMB dipole prediction is also removed from the timelines, including both the orbital dipole and the solar component, evaluated 
following the methodology outlined in Section 4.4.2 of \cite{picasso}. Estimates for the RFI signals are removed at the CTOD level in the MFI pipeline using a procedure based on azimuthal stacks \citep[see Sec.~2.2.3 of][]{mfiwidesurvey}. Templates for these signals are computed over entire observing periods lasting several months, and are generated separately for each elevation. This method relies on the assumption that the RFI signal remains stable in local coordinates throughout the entire period.

We note that MFI wide survey CTOD files already incorporate some data flagging \citep[see Sec.~2.2.2 in][]{mfiwidesurvey}, including the gaps associated with the use of the calibration diode (1\,s every 30\,s), and multiple flags due to voltage ranges, house-keeping parameters, emission of the Sun and the Moon, emission of geostationary satellites, and a specific flagging based on the root-mean-square of the data in each scan.

It is important to note that the QUIJOTE MFI instrument is not sensitive to the mean (zero-level) sky emission. The median sky signal, comprising contributions from the CMB monopole and the average atmospheric emission, is subtracted from each azimuth scan. However, spatial variations in PWV during the telescope’s motion can introduce large-scale patterns in the timelines, which are subsequently projected onto the sky during the map-making process. 

\begin{figure*}
    \centering
    \includegraphics[width = 18cm]{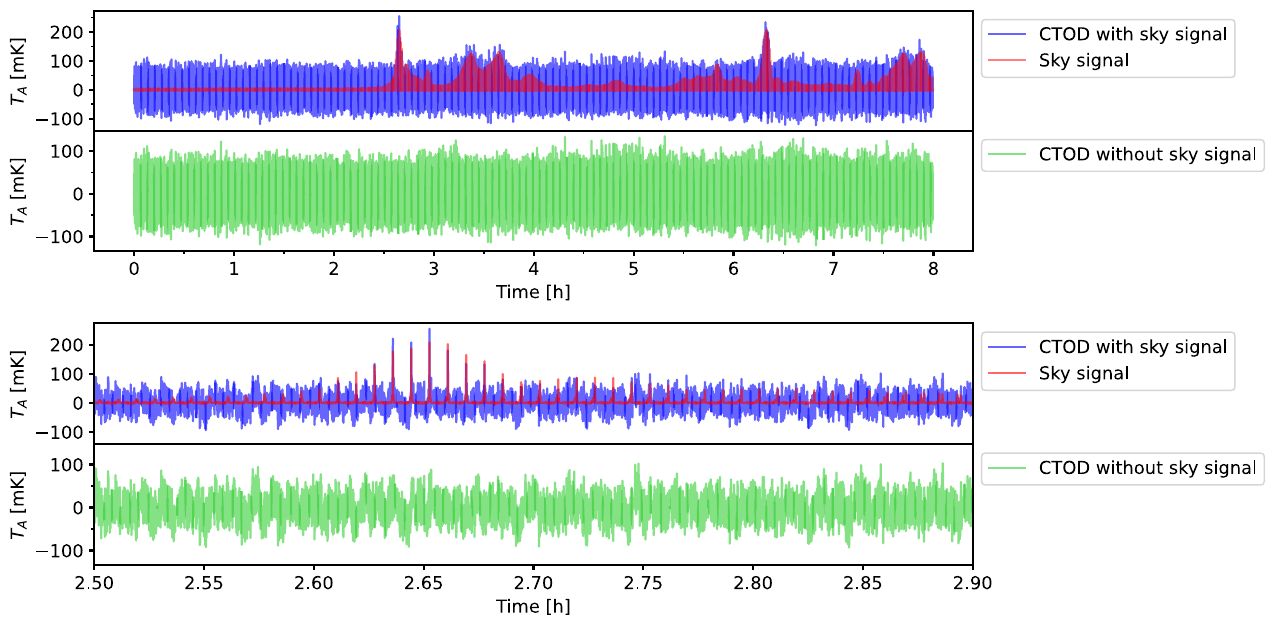}
    \caption{ Top: example of a calibrated timeline for an 8-hour observation from the QUIJOTE MFI wide survey (antenna temperature $T_\text{A}$ against time). This observation was recorded on the 11th of January 2017, at 08:35 UTC+0, with the telescope pointing at an elevation of $60^\circ$. Potential RFI signals were removed or flagged by the MFI pipeline. The blue curve shows the total intensity signal measured by horn 2 at 17\,GHz before removing the astrophysical sky and dipole signals. The expected astrophysical sky signal is shown in red. The green curve shows the same signal after removal of the expected sky contribution. Bottom: zoomed signals between 2.5 and 2.9 hours before and after the sky signal correction. The intense peaks visible around 2.65 hours are due to the crossing of the galactic plane (at a position with galactic longitude of $49.26$°). 
    %Additional peaks arise from the telescope’s scanning strategy.
    }
    \label{fig:time_sig}
\end{figure*}

Figure~\ref{fig:time_sig} illustrates this process of signal removal. The blue curve in the top panel shows the intensity signal of one observation of the MFI wide survey before the removal of the astrophysical sky signal and the CMB dipole, and the green curve in the top panel shows it after that step. The total projected astrophysical sky signal is shown in red in both panels. On this figure, the intensity ranges from $-100$ to $200$\,mK. The bottom panel shows the same zoomed in around the first peak in the sky signal.
%This is due to the fact that the median intensity from each 30-second telescope's scan was subtracted. %This is standard procedure allowing for the suppression of the very long-term intensity variation, such as the one induced by the atmospheric signal. 
%For each MFI observation, a set of CTODs was produced in which the median intensity of each 30-second telescope scan was subtracted. 
%Moreover, flagged samples can be seen in this observation as intensity 'gaps', i.e. intensity values set to 0\,mK and are visible at 1.1 and 2.8 hours. %We proceeded similarly to remove the CMB dipole at the CTOD level. 
%: we projected a simulated CMB dipole map into the horn CTOD to obtain the contribution from the dipole and then removed it. 

%We investigate both contributions in the next subsections. 

%--------------------------------------------------------------------
\subsection{Atmospheric signal in the QUIJOTE MFI wide survey}

\begin{figure*}
    \centering
    \includegraphics[width = \textwidth]{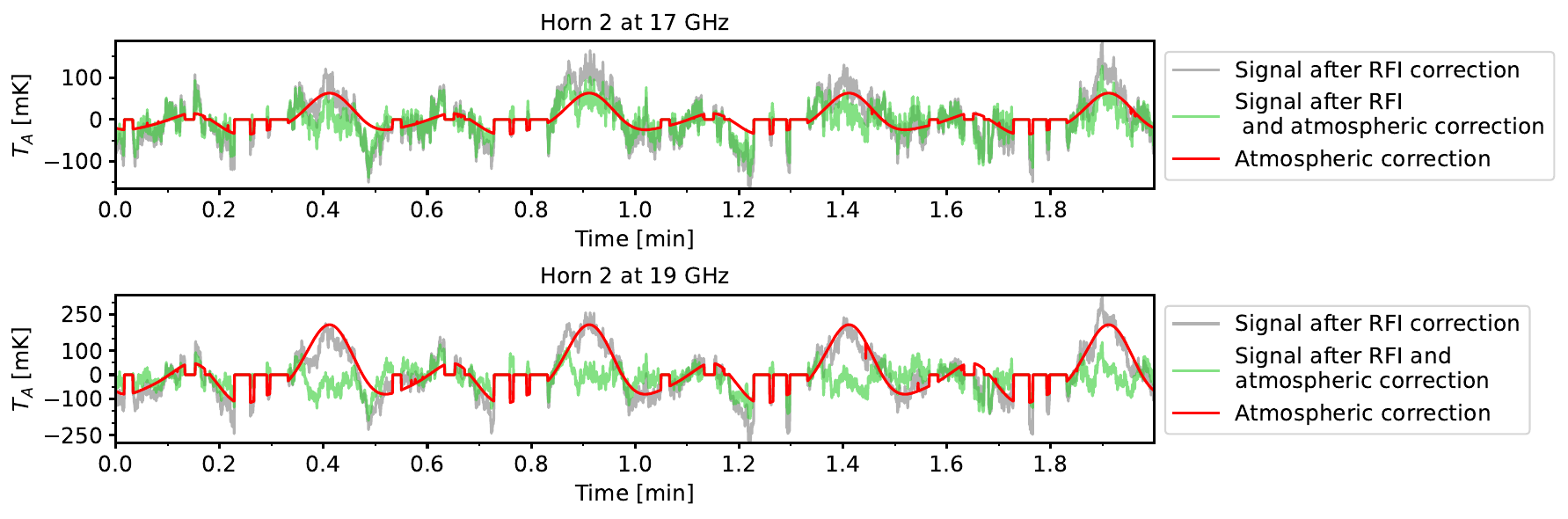}
    \caption{Illustration of the effect of atmospheric correction through the PCA analysis at the TOD level. We show measured calibrated antenna temperature $T_\text{A}$ as a function of time for two minutes of a QUIJOTE MFI wide survey observation taken on August 30th 2016 at 14:45 UTC+0, and at an elevation of $50^\circ$. The median PWV during the CTOD acquisition was 14\,mm, which is very high and should produce a strong atmospheric signal. The grey curve represents the data after the application of RFI correction; the red curve is the atmospheric signal obtained with PCA, and the green curve represents the data after RFI and atmospheric corrections. The expected astrophysical signal and CMB dipole were removed from the data, so the residuals should be consistent with noise. }
    \label{fig:signals_comp}
\end{figure*} % NOMINAL50A-160830-1445-000

%The frequency bands of QUIJOTE MFI are contaminated by atmospheric emission, mainly arising from the emission line of water vapour at 22\,GHz, as seen from Figure \ref{fig:TB_atmo}. These emissions contribute both to the average sky load of the detectors, but also create noise due to their stochastic nature. 

An atmospheric correction was implemented in the pipeline of the QUIJOTE MFI wide survey to mitigate the impact of atmospheric emission on the final intensity maps. Before the map-making step, broad-scale features with frequency dependence across the four MFI bands consistent with atmospheric emission were removed from the data \citep[see Sec.~2.2.4 of][]{mfiwidesurvey}. These features were identified using Principal Component Analysis (PCA) applied to one-hour-long azimuthal stacks. The choice of a one-hour duration represented a compromise between achieving a decent signal-to-noise and avoiding the effect of time variations of the atmosphere. Thus, this method relied on the assumption that atmospheric conditions remain stable over a one-hour period. This assumption will be tested in this paper.
 
Figure~\ref{fig:signals_comp} illustrates the atmospheric correction for a single MFI wide survey observation. We show 2 minutes of data as measured by horn 2 at 17\,GHz (top) and 19\,GHz (bottom), after subtracting the median intensity from each 30-second telescope scan. The grey curve shows the calibrated sky intensity emission after RFI correction but before atmospheric correction. 
The red curve shows the atmospheric signal obtained using the PCA method described above, and the green curve shows the final signal after both atmospheric and RFI corrections.
Several effects can be observed in this figure. First, the amplitude of the PCA correction confirms that the large-scale atmospheric signal associated with spatially varying PWV conditions is the dominant source of emission at 17 and 19\,GHz in this observation, which was taken on a day of particularly high PWV conditions (14\,mm). The atmospheric emission in this case exceeds the CMB dipole by one to two orders of magnitude. As expected, its amplitude is stronger at 19\,GHz than at 17\,GHz due to the proximity of the 22\,GHz water vapour line. It can also be noted that at 17\,GHz the amplitude of the atmospheric emission seems a bit low compared to the data, the reason being that this plot shows just 2\,min of data while the fit is performed on 1\,h of data.
%Second, we observe a near-sinusoidal atmospheric pattern caused by varying weather conditions across different azimuth angles as the telescope rotates to scan the sky in circles, as explained in \ref{section:QUIJOTE_MFI2}. 
Second, the green curve reveals residual noise, even after the removal of RFI and large-scale atmospheric contributions. This remaining noise is probably a combination of low-level atmospheric fluctuations and instrumental components, both exhibiting a $1/f$ behaviour.

%
%Overall, this figure, along with Figure \ref{fig:time_sig}, demonstrates that atmospheric contamination dominates the intensity signal at both frequencies, exceeding the CMB emission by one to two orders of magnitude. This also confirms that, after subtracting the astrophysical sky signal and the dipole, the remaining signal is due to the atmosphere and not some leftover contribution from another source, confirming that our analysis is indeed characterising the atmospheric fluctuations.

%Spatial variations in PWV across the sky lead to significant variations in the measured atmospheric temperature. 
%The atmosphere acts as a radiating medium, emitting approximately as a blackbody at the physical temperature $T_\text{atm}$. Solving the radiative transfer equation yields an expression for the antenna temperature $T_\text{A}$ as a function of  airmass $m$, the atmospheric physical temperature and the opacity $\tau$:
%
%\begin{equation}
%    T_\text{A} = T_\text{atm} (1 - e^{-\tau\cdot m(\theta)})\approx  T_\text{atm} \tau \cdot m(\theta) = T_\text{A}^\text{z} \cdot m(\theta),
%\end{equation}
%
%being $\theta$ the telescope elevation, $m(\theta)$ the airmass (approximated here as $m(\theta) = 1/\text{sin}(\theta)$), and $T_\text{A}^\text{z}$ the antenna temperature at zenith. The approximation included in that equation is done for the optically thin regime ($\tau \ll 1$). 
%
%
%With this formalism, antenna temperature variations can be now converted into PWV variations. 
%

Finally, we can translate the measured antenna temperature variations into spatial variations of PWV. For instance, using Figure~\ref{fig:TB_atmo}, we obtain that at 19\,GHz, a change in PWV of 1\,mm corresponds to a change in atmospheric brightness temperature of $\Delta T_\text{A}^\text{z} = 0.443 \text{K}$, equivalent to a change in the antenna temperature at elevation $50^\circ$ of $\Delta T_\text{A}= 0.443$\,K $\cdot m(50^\circ) = 0.578$\,K. Therefore, a 1\,mm variation in the PWV condition across the sky can produce changes in the measured temperature as large as 0.6\,K for a telescope elevation of $50^\circ$, consistent with the amplitude of the signals seen in Figure~\ref{fig:signals_comp}.

%\noindent
%The Von Kármán spectrum \citep{vonkarman1948, goedecke2004} is used for $\kappa < 1/L_0$, where the assumptions of homogeneity and isotropy
%of the atmospheric turbulence is usually invalid.

%--------------------------------------------------------------------
\section{Time stability of the atmospheric signal in QUIJOTE-MFI data} \label{section:cross_corr} \label{section:CL}

% NOMINAL60C-170111-0835-000.ctod
% 'NOMINAL60D-170107-1700-000.ctod'

%[THIS PARAGRAPH COMES FROM SECT. 2. TO BE INTEGRATED]
%In the second part of this paper (see Section \ref{section:CL}), we analyzed the cross-correlation function of the signal of horn 2 and horn 4 at 17\,GHz and 19\,GHz using MFI wide survey observations to analyse the correlation of the atmospheric signal across different horns. For this analysis, the number of flagged samples (the flagging of QUIJOTE MFI is described in \citet{mfiwidesurvey}) in each observation was crucial because it significantly affects the correlation function. Flagged samples were set to zero, which reduces the overall correlation by artificially lowering the product of data pairs. As a result, a higher proportion of zeros biases the correlation function downward, making it appear weaker than it truly is. Therefore, it was essential to exclude from the analysis the observations containing too many flagged samples. After analysing the impact of the flagging on the cross-correlation function, we decided to keep files containing less than 30\% of flagged samples to compute the cross-correlation function, as the effects below this value were negligible. We therefore removed 711 observations from the 1233 MFI observations on this flagging condition for this analysis, which is about 58\% of the observations. More details about this data selection are given in the Appendix \ref{section:data_sel}

In this section, we test the assumption on the time stability of the atmospheric signal over 1-hour periods used in the QUIJOTE MFI wide survey papers. 
%the atmospheric contamination was corrected every hour. This means that the data were divided into 1-hour segments, and a Principal Component Analysis (PCA) was applied to each segment to extract the atmospheric signal. The estimated atmospheric contributions were then subtracted from each of their corresponding 1-hour data segment. The assumption made was hence that the atmospheric signal stays identical over a 1-hour period \citep{mfiwidesurvey}. This time of one hour was a trade-off between the varying integrated water vapour content along the line of sight as the weather changes over the observatory, which is expected to change completely in a range of several hours, and the signal-to-noise ratio of the PCA analysis.
%
%In order to verify the assumption, 
To that aim, we have performed a statistical analysis of the intensity signals of QUIJOTE MFI horns 2 and 4 data to measure the coherence length in time of the correlated atmospheric signal. After subtracting the astrophysical sky signal, dipole signal and RFI from each horn's output, the only common remaining component should be the atmospheric signal plus the correlated noise component. As described earlier, each horn has independent instrumental noise, so their noise contributions are uncorrelated.  Hence, if we compute the cross-correlation of the two horns, we will obtain the correlation due to the atmospheric signal. It is important to note that the angular separation on the sky of the pointing of horns 2 and 4 is $5.6^\circ$, and then the cross-correlation of the data from these two horns gives the correlated atmospheric signal on this angular scale.
%Since the atmospheric signal is expected to stay identical over a period of 1 hour, the cross-correlation of both horns' signals is also expected to be present for such a period. 
%The goal of the analysis was to determine the temporal stability of the atmospheric signal's cross-correlation across different sky directions during the MFI wide survey.

We calculated the cross-correlation function at 17\,GHz and 19\,GHz using wide survey data in which the telescope scans the sky in azimuth circles of $360^\circ$ at a constant elevation. In order to evaluate the time stability of a given area on sky, as measured in local coordinates, we selected samples with the same azimuth. 
%In this way, even if the telescope was moving, we removed the effect of the spatial variation of the atmosphere and are only sensitive to the time variation. 
As the scanning speed of the QUIJOTE telescope during the wide survey is 12\,deg/s, we are not sensitive to time variations on time scales shorter than 30 seconds, which is the time duration of one full scan. 
For each azimuth value considered, we then extract 1 sample per telescope scan, building CTOD of samples of the same azimuth with a sampling frequency of 30\,s. We covered azimuths from $0$ to $350^\circ$ in increments of $10^\circ$.

%The cross-correlation function removes instrumental noise from each horn, as the noises from different horns are uncorrelated. This is because each of the four QUIJOTE MFI horns has its low-noise amplifier (LNA), generating independent instrumental noise signals.

%If one computes the autocorrelation function of the signal of the same horn, the instrumental noise will contribute to the correlation. On the contrary, computing the cross-correlation function of different horns allows us to compare the only signal present in both horns, i.e. the atmospheric signal.

The discrete (normalised) cross-correlation function $C_{d_2 d_4}(\tau)$ of the two horn data vectors $d_2$ and $d_4$ at a time lag $\tau$, which is a discrete variable with step 30\,s, was calculated according to
\begin{equation} \label{eq:corr_fct}
    C_{d_2 d_4}(\tau) = \frac{\sum_{n=0}^{N-1} \overline{d_2(n)} d_4(n + \tau)}{ \sqrt{\sigma_{d_2}^2 \cdot \sigma_{d_4}^2}},
\end{equation}
being $N$ the number of samples in the two data vectors $d_2$ and $d_4$, $\overline{d(n)}$ the complex conjugate, and $\sigma_{d_2}^2$ and $\sigma_{d_4}^2$ the variance of the signals (or auto-correlation function at lag zero), which was used to normalise the cross-correlation function.
Our numerical implementation makes use of the {\verb!correlate!} package from {\sc SciPy}. 
%The normalisation allowed us to characterise the amount of correlation more easily. 
The application of normalisation in equation~\ref{eq:corr_fct} results in the measurement of quantity being the relative amount of correlated signal, a quantity that is easier to interpret.
%For example, a normalised correlated peak with an amplitude of 0.6 indicates that 60\% of correlation in the signals. 
However, an important effect to consider is that, at lag 0, the autocorrelation includes both the atmospheric signal and instrument noise. 
%Since the noise is random and uncorrelated between horns, it does not appear in the cross-correlation. 
As a result, the normalisation factor is overestimated due to this noise contribution, which causes, in turn, the cross-correlation amplitude to be underestimated. This explains why the maximum values of the correlated signals are always lower than 1 on those cross-correlation plots.  In practice, the closer the amplitude to one, the higher the signal-to-noise ratio of the atmospheric signal in that particular azimuth direction.

\begin{figure*}
    \centering
    \includegraphics[width = 18cm]{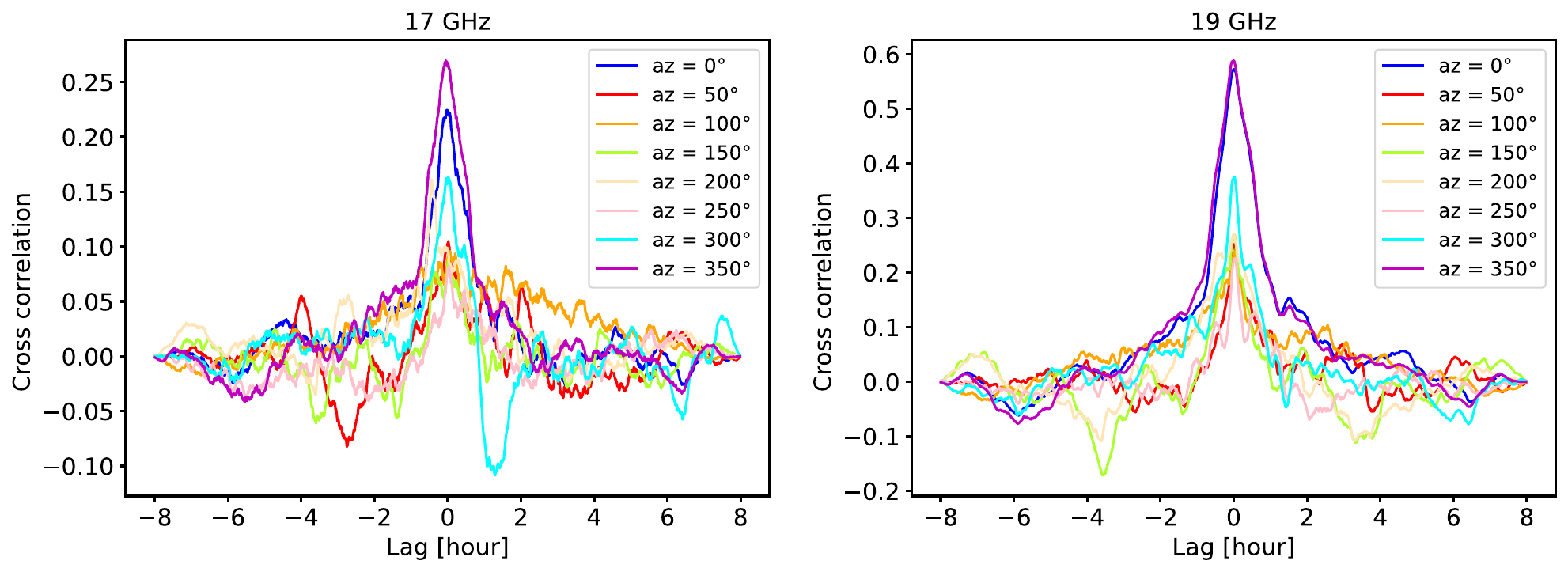}
    \caption{Normalised cross correlation functions of horn 2 and horn 4 for one observation of the QUIJOTE MFI wide survey at different azimuths at 17\,GHz (left) and at 19\,GHz (right) calculated according to equation \ref{eq:corr_fct}. This observation was made on the 11th of January 2017 at 8:35 UTC+0. The PWV was 1.9\,mm. The median wind speed was 2.4\,m/s and the median wind direction was 325\,° (North-West) at the observation time. The cross-correlation functions are computed for samples selected at specific azimuths, covering azimuths from 0° to 350° in 50° increments, to compare different regions of the sky. Moreover, the curves are smoothed using a kernel of 20 samples to reduce the impact of noise.}
    \label{fig:cross_corr_all_az}
\end{figure*}

%Moreover, for the calculation of the cross correlation function, we didn't introduce time shifts but shifts by $N$ samples, the time separation between two samples of one scan being 40\,ms, and the time between samples of the same azimuth being 30\,s.

% auto-correlation functions at a lag of 1 sample (30\,s), because the instrument white noise signal dominates the auto-correlation function at lag 0 sample

%However, an effect to take into account is that, at lag 0, the auto-correlation includes both the atmospheric signal and the noise of the instrument. The normalisation factor obtained with those auto correlation values is therefore inflated by the noise, which is not included in the cross-correlation, because noise is random and doesn’t correlate between horns. Consequently, using this normalisation underestimates the cross-correlation amplitude, since it will reflect both the atmospheric contribution and a minor contribution from the noise.

After selecting all samples at a given azimuth value, the correlation functions between horns 2 and 4 are computed for all possible time lags and plotted against the lag. 
%A maximum in the correlation function implies a strong similarity of the signals, while a correlation curve close to zero indicates no correlation. 
On those plots, we typically observe a correlation peak at zero lag. This peak indicates that the two horns are seeing a 
%correlated signal. Our interpretation is that this correlation is sourced by a 
common atmospheric signal. We can retrieve several pieces of information from the peak. First, from the amplitude of the peak, we can extract the degree of correlation between the two signals. Second, from the width of the peak, we can extract the time during which those signals remain correlated. This will tell us how long the atmospheric emission remains similar.
%, telling us about the rate of change of the PWV as well. 

Figure~\ref{fig:cross_corr_all_az} shows the results for one observation of QUIJOTE MFI at different azimuths in $50^\circ$ steps at 17\,GHz (left) and 19\,GHz (right). For this observation, the maximum correlation of the signals reaches about 57\,\% at azimuth $350^\circ$ (north-east) at 19\,GHz and about 27\,\% at 17\,GHz in the same direction. However, it is important to note that the curves have been smoothed over 20 data points (equivalent to 10 minutes), a process that might reduce the maximum amplitude of the cross-correlation peak. The lowest correlation amplitudes are found between azimuth $50^\circ$ (north-east) and $250^\circ$ (south-west). We can conclude from this plot that the atmosphere was low on PWV in those directions on that day, consistent with the median PWV of 1.9\,mm recorded during the observation by the GPS antenna. More generally, we can conclude that the atmospheric signal is strongly dependent on the direction of the sky. This highlights the highly uneven distribution of water vapour in the atmosphere, such as that caused by the presence of clouds. Another effect that can be highlighted from this observation is that the correlation is lower at 17\,GHz than at 19\,GHz. This is again explained by the proximity of water emission lines at 22\,GHz (Figure~\ref{fig:TB_atmo}).

\begin{figure}
    \centering
    \includegraphics[width = \columnwidth]{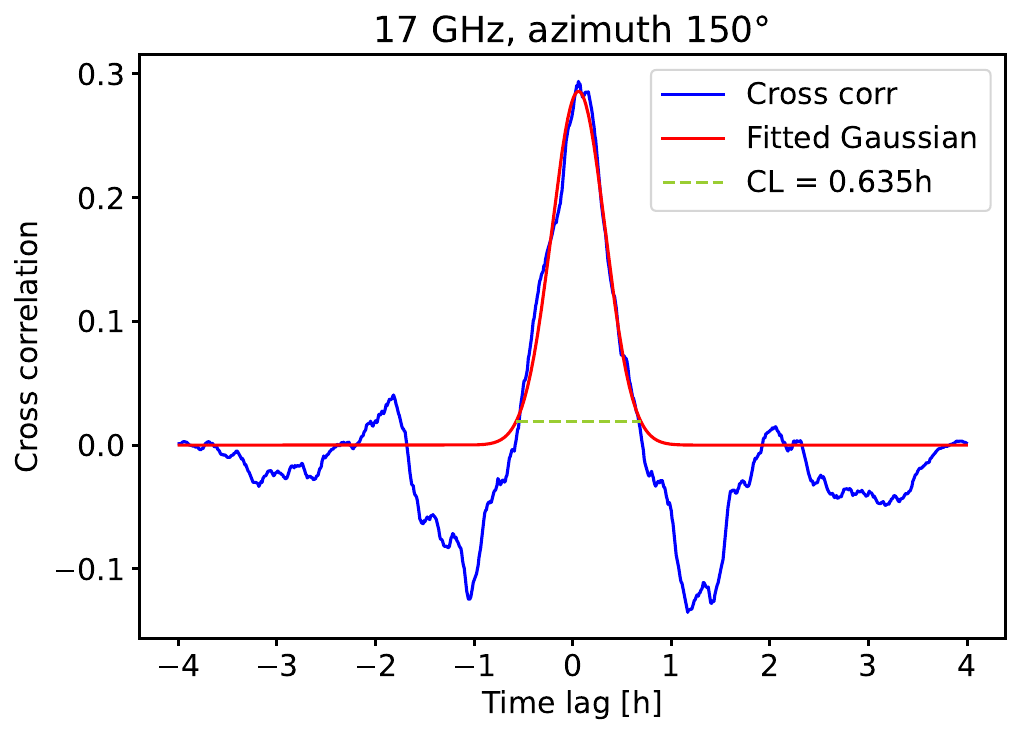}
    \caption{Normalised cross correlation function of horns 4 and horn 2 at 17\,GHz and azimuth $150^\circ$ smoothed over 20 data points for an observation of the QUIJOTE MFI survey, computed according to equation \ref{eq:corr_fct}. This observation was taken on the 22th of November 2013 at 9:47 UTC+0. The PWV was 7.4\,mm. The cross-correlation is represented in blue. The cross-correlation peak is fitted with a Gaussian curve (red line) to extract the values of the CL (green dashed line). }
    \label{fig:CL_fit_ex}
\end{figure}

% 17GHz: CC_17_az150_NOMINAL60B-131122-0947-000.ctod.pdf

To quantify the time coherence of the correlation between the two signals, we define the Coherence Length (CL) as %the characteristic time over which the signals remain correlated. This quantity gives the duration of stability of the atmospheric signal. The coherence length was determined from 
half the width of the correlation peak at $1/15$ of the maximum amplitude. Operationally, this is obtained by fitting a Gaussian curve to each dataset's correlation peak. 
%and by extracting the curve's width at $1/15$ of the Gaussian maximum amplitude, and then dividing it by 2. 
We took the half-width because this is what represents the typical time scale in which the atmospheric signal is correlated.
%The true time over which the atmosphere is correlated is then the half-width. 
%Mathematically, this is expressed as 
%\begin{equation}\label{eq:CL}
%   CL =  \sqrt{2  \ln(15)} \cdot \sigma ,  
%\end{equation}
%
%\noindent
%with $\sigma$ the standard deviation of the Gaussian fit. 
%We chose to select the width of the peak at an amplitude of 1/15 of the Gaussian maximum as a compromise to avoid underestimating or overestimating the correlation peak's width. 
%This threshold was selected to approximate the point where the correlation reaches zero, accounting for the fact that the Gaussian fit never reaches 0.

We have used this Gaussian fitting methodology for several reasons. First, fitting a Gaussian curve instead of measuring the correlation peak directly allows for reducing the effect of the noise. Noise causes bumps and irregularities in the cross-correlation, making it hard to accurately identify the true width of the correlation peak. Our method thus helps to improve the stability and robustness of the measurement. Second, this method allowed for the fully automated extraction of the CL over the entire MFI wide survey observations. Given the large number of observations, this automation was essential to efficiently process the data and minimise human bias. On the other hand, the downside of this method is that the peak may not be a perfect Gaussian function.
%and, as a result, the width of the bottom of the Gaussian function might deviate from the actual correlation peak. 
Figure \ref{fig:CL_fit_ex} illustrates the Gaussian fitting process for an observation of the MFI survey, at 17\,GHz and azimuth $150^\circ$. In that example, the cross-correlation function is represented in blue, the Gaussian fit in red, and the extracted CL, of about  $0.64$\,h, in green. 
%However, the green dashed line shows the full width of the peak, but the CL is actually half that width.

\begin{figure}
    \centering
    \includegraphics[width = \columnwidth]{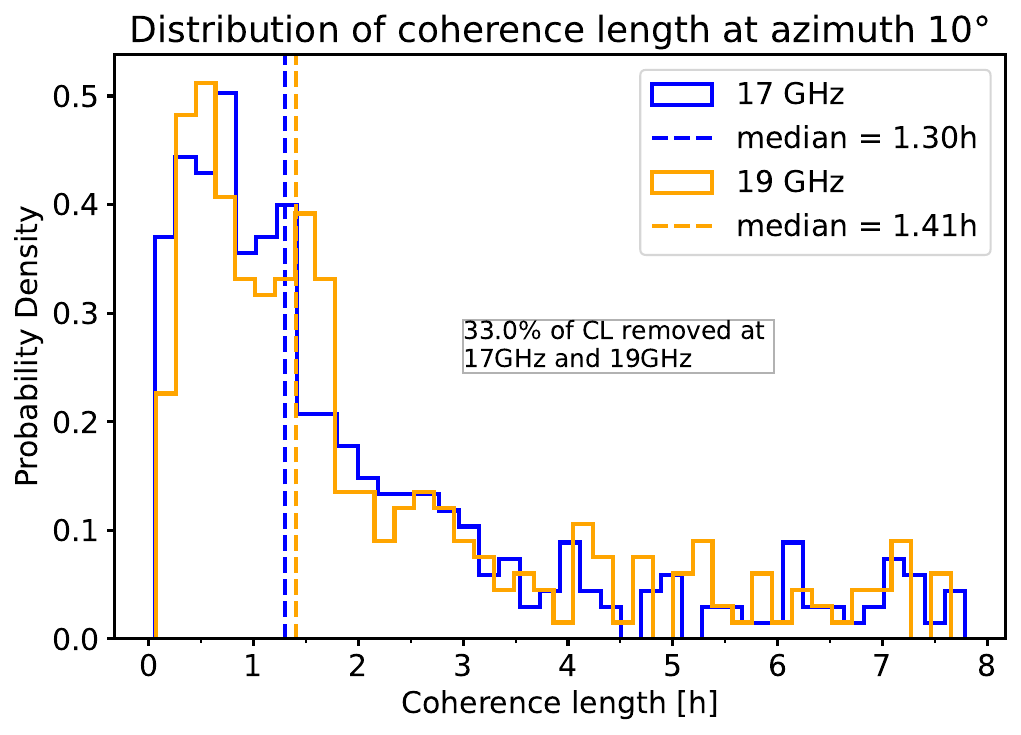}
    \caption{Distribution the coherence length results obtained at azimuth 10° for each selected observations at 17\,GHz (blue) and 19\,GHz (orange). For this analysis, the observations with coherence lengths set to zero were removed according to the above conditions. The dashed lines are the medians of the distribution at both frequencies.}
    \label{fig:hist_CL}
\end{figure}

%We extracted the CL for the observations of the wide survey covering azimuths from 0 to $360^\circ$ in $10^\circ$ steps to obtain a full azimuth coverage. 
For some of the 1223 QUIJOTE MFI wide survey observations it was not possible to extract the CL measurement, because of various reasons. First, we removed observations with more than 30\,\% of flagged samples in their intensity signal, as very severe flagging might bias the determination of the CL. In total, we excluded 711 observations based on this condition.
Second, we removed observations for which the fitting procedure was not converging, mainly due to the limited number of samples. Observations with less than 25.000 samples (equivalent to a time duration of 16 minutes) were removed. If a criterion failed for a given observation at a specific azimuth, either at 17\,GHz or 19\,GHz, the CL was set to zero for both frequencies at that azimuth. This ensured that we obtained the same number of CL measurements across the two frequency bands. Finally, all observations with a CL set to zero were excluded from the analysis. 
The median number of observations kept at each azimuth is hence 337. Section~\ref{section:data_sel} gives more details on the data selection process, including the exact number of observations used at each azimuth.

Figure~\ref{fig:hist_CL} shows the distribution of coherence lengths (CLs) resulting from 367 observations at 17\,GHz (blue) and 19\,GHz (orange) at azimuth $10^\circ$. 
%Approximately 33\,\% of the observations were excluded due to poor fitting quality or insufficient samples. 
The distribution peaks at around 0.8~hours. However, most observations have CLs exceeding 0.8 hours, with median values of 1.3 hours at 17\,GHz (blue dashed line) and 1.4 hours at 19\,GHz (orange dashed line). Beyond 3 hours, the distribution drops sharply, indicating a rapid decline in the probability density as CL increases. There are, however, a few observations with very high CL values, with the maximum reaching 7.6 hours. Overall, the CL results span a wide range, indicating a variety of atmospheric conditions—ranging from highly stable scenarios (steady wind direction and speed) to rapidly varying ones characterised by high wind speeds and frequent directional changes. The distribution at this azimuth shows the general pattern seen at other azimuths, although the median values vary, as shown below.

\begin{figure}
    \centering
    \includegraphics[width = 9cm]{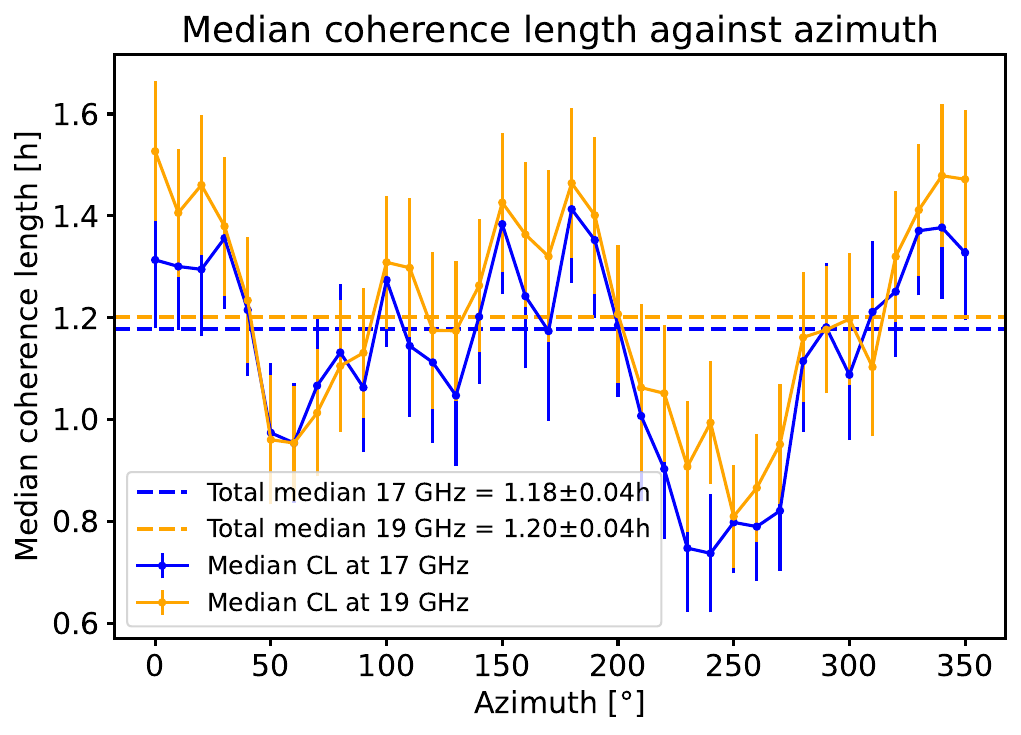}
    \caption{Median coherence length as a function of azimuth, with error bars at 17\,GHz (blue) and 19\,GHz (orange) from the QUIJOTE MFI selected observations. Dashed lines indicate the overall median coherence lengths at each frequency.}
    \label{fig:CL_vz_az}
\end{figure}

 The median CLs obtained at each selected azimuth with error bars are shown in Figure \ref{fig:CL_vz_az}. The median CLs at both 17 and 19\,GHz of the selected observations were obtained at each selected azimuth $i$. The error bars $\Delta \text{CL}_{i}$ were calculated according to the Standard Error of the Mean (SEM). The overall median is 1.18 hours at 17\,GHz and 1.20 hours at 19\,GHz. This difference is small compared to the error bars. This is expected given that the coherence length should in principle be the same at both 17 and 19\,GHz.
 The median CLs as a function of azimuth trace a similar pattern at both frequencies, and their values generally fall within each other’s error margins for almost all azimuths. Furthermore, there are variations in the median CLs depending on azimuth, especially at 50\,° (north-east) and 250\,° (south-west), where the CLs are much lower, with values of the order of 0.8 hours. It is not clear what could be the cause of this, but we hypothesise that the wind direction or the topology of the Teide Observatory could have an effect. In general, we demonstrated that the atmospheric signal is stable on time scales of $\sim $1 hour, confirming the validity of the atmospheric signal. PCA correction of the MFI wide survey data. 

% This might arise from a bias in the analysis where the atmospheric signal is stronger at 19\,GHz, which may lead to a higher correlation peak and, thus, a broader peak.

%This might result from the wind direction since the wind blows from the north-east and might move the atmospheric turbulence more quickly in front of the telescope at those azimuths.

%This result shows that experimental results agree with theoretical expectations. This consistency supports the underlying hypothesis in this analysis, that all the correlated signals are due to atmospheric emission, and is verified. 
\begin{figure}
    \centering
    \includegraphics[width=9cm]{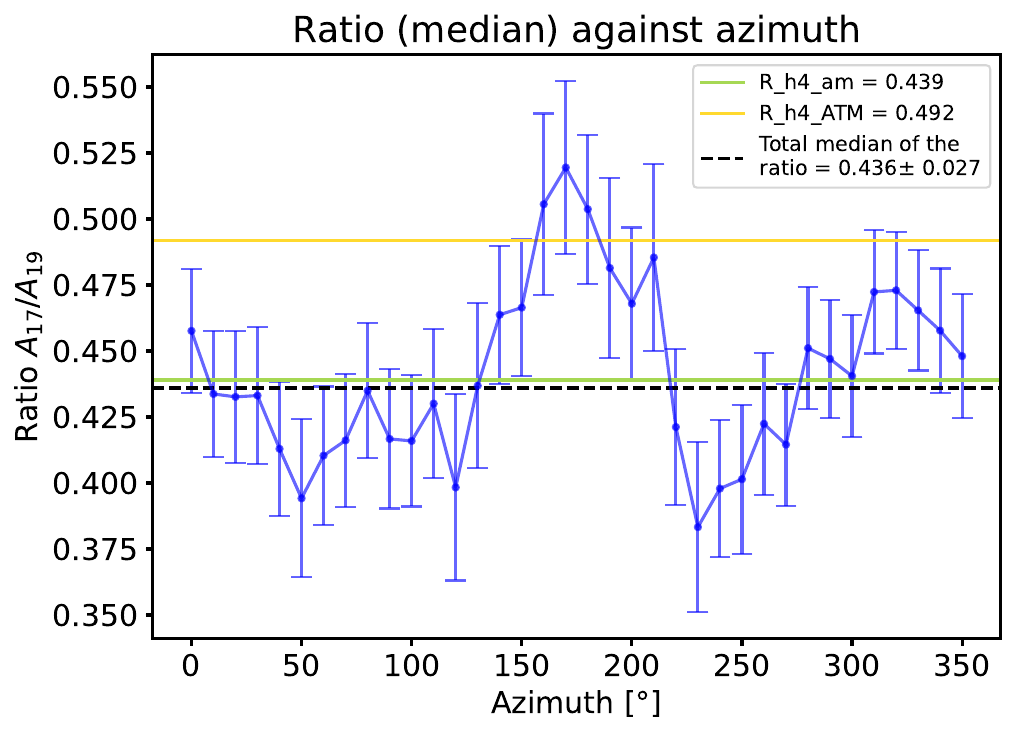}
    \caption{Median ratio of the amplitude of the cross-correlation of horn 2 and 4 at 17 GHz and the amplitude of the cross-correlation of horn 2 at 17 GHz and horn 4 at 19 GHz, according to Equation \ref{eq:ratio_exp}. The theoretical ratios obtained with {\sc am} for horn 2 and horn 4 are shown in light green and deeper green, respectively. The ones obtained from {\sc ATM} are shown in orange for horn 2 and in yellow for horn 4. The median ratio across azimuth is shown as a dashed black line.}
    \label{fig:median_ratio}
\end{figure}

%The fact that we get consistency with the models gives confidence in that initial assumption.

\subsection{Ratios of the correlated signals at 17 and 19\,GHz} 
\label{section:ratio}

The unnormalized amplitudes of the cross-correlation signals at different frequencies can be used to test the validity of the models that describe the mean atmospheric emission ({\sc am} and {\sc atm}) as follows.
%
%The next step in the analysis was to compare our results with the theory. 
%To do that, we measure the ratio between the amplitudes of the cross-correlation peaks at 17 and 19\,GHz. We can then compare this ratio with the theoretical ratio computed with {\sc am} \citep{Paine} and {\sc ATM} \citep{ATM}. 
Our reference observable will be the measured ratio between the amplitudes of the cross-correlation peaks at 17 and 19\,GHz. 
To obtain the theoretical value for that ratio, we 
can use the brightness temperature curves
at the OT for several PWV contents, computed both with {\sc am} and {\sc atm} (see e.g. Figure~\ref{fig:TB_atmo}).

%of the atmospheric profile across the four seasons at OT for several PWV contents with {\sc am} and {\sc ATM} as seen in Figure \ref{fig:TB_atmo}, top panel for {\sc am}. 
To account for the MFI spectral response, we estimate the effective antenna temperature, in Rayleigh-Jeans brightness temperature units, by computing the bandpass-weighted frequency integral of the model output (the bandpasses of horns 2 and 4 of QUIJOTE MFI are shown in Figure \ref{fig:TB_atmo}, bottom panel). 
%This band-averaged temperature takes into account the frequency-dependent sensitivity of the receiver. 
This is expressed mathematically as

%Then, for each PWV content, we took the value of the atmospheric emission model $T_\text{RJ}$ and integrated it over the instrument’s frequency response, i.e. the bandpass of QUIJOTE horn 2 and 4 shown in Figure \ref{fig:TB_atmo}, bottom panel.

\begin{equation} \label{eq:integral}
    T_\text{A}^\text{eff} = \frac{\int T_\text{RJ}(\nu) g(\nu) \text{d}\nu}{\int g(\nu) \text{d}\nu},
\end{equation}

\noindent
with $g(\nu)$ the bandpass, $\nu$ the frequency, and $T_\text{RJ}(\nu)$ the Rayleigh-Jeans brightness temperature of the atmosphere given by the model.

\begin{table}
    \centering
    \begin{tabular}{ccccc }
        & \multicolumn{2}{c
        }{a} & \multicolumn{2}{c}{b} \\ \cline{2-5}
        & {\sc am} &  {\sc ATM} & {\sc am} & {\sc ATM} \\ \hline \hline
        Horn 2, 17\,GHz & 0.210 & 0.262 & 1.824 & 1.865 \\
        Horn 2, 19\,GHz & 0.485 & 0.536 & 2.006 & 2.134 \\
        Horn 4, 17\,GHz & 0.219 & 0.269 & 1.833 & 1.877 \\
        Horn 4, 19\,GHz & 0.498 & 0.548& 2.011 & 2.146 \\ \hline \hline 
        \end{tabular}
    \caption{Fitted parameters obtained from {\sc am} and {\sc ATM} model of the atmospheric temperature observed by QUIJOTE horns 2 and 4 at 17 and 19 GHz as a function of PWV, derived from Equation \ref{eq:fit_poly}. These fitted functions will be used to compute the theoretical amplitude ratios.}
    \label{tab:coeff}
\end{table}

We applied equation~\ref{eq:integral} to the bandpasses of horns 2 and 4 at 17 and 19\,GHz, for the simulated brightness temperature for each PWV value (23 different values ranging from 0.1 to 25\,mm for {\sc am} and 27 different values ranging from 0.1 and 50\,mm for {\sc atm}). We hence obtain 23 values of the effective antenna temperature $T_\text{A}^\text{eff}$ for {\sc am} and 27 for {\sc atm}. We then plotted those values against their corresponding PWV. This resulted in a relationship that can be fitted with a linear function as:
\begin{equation}\label{eq:fit_poly}
T_\text{A}^\text{eff}(PWV) = a \cdot \text{PWV} + b.
\end{equation}
The intercepts $b$ represent the combined contributions of $O_2$ and the CMB monopole, which are not measured by QUIJOTE MFI as they contribute as a constant baseline level at a given elevation. 

The results of these fits are shown in Table~\ref{tab:coeff}. The coefficients $a$ are slightly higher for {\sc atm} as compared to {\sc am}, leading to differences in the resulting theoretical predictions. These discrepancies stem from differences in the atmospheric modelling implemented in these packages. However, it is important to note that the opacity predictions from the {\sc am} model generally align better with what we measure with QUIJOTE than those from {\sc ATM} \citep{Chappard2025}. This is in some way expected as {\sc am} uses specific temperature and pressure profiles for each site, whereas {\sc atm} relies on a reference constant profile. Therefore, we consider {\sc am} predictions more reliable.

The expected theoretical ratio $R_\text{theo}$ of the brightness temperature at 17 and 19\,GHz can then be obtained from the above coefficients for {\sc am} and {\sc atm} as:
\begin{equation} \label{eq:ratio_am}
    R_\text{theo} = \frac{\Delta T_\text{A 17}^\text{Eff}}{\Delta T_\text{A 19}^\text{Eff}} =  \frac{a_{17}}{a_{19}}.
\end{equation}
%The obtained ratios are shown in Figure \ref{fig:median_ratio} in green shades for {\sc am} and in orange shades for {\sc ATM}. 

The experimental ratio $R_\text{exp}$ can be obtained from the unnormalized cross-correlation function of signals of horn 2 and 4 at 17\,GHz divided by the cross-correlation function of signals of horn 2 at 17\,GHz and horn 4 at 19\,GHz. This ratio can be directly compared to the theoretical ratio, since we can write:
\begin{align} \label{eq:ratio_exp}
    R_\text{exp} &= \frac{<T_{2,17}, T_{4, 17}>}{<T_{2,17}, T_{4, 19}>} \\ \nonumber
    &= \frac{ <a_{2,17} \cdot \text{PWV} , a_{4,17} \cdot \text{PWV}>}{ <a_{2,17} \cdot \text{PWV} , a_{4,19} \cdot \text{PWV}>} \\ \nonumber
    &= \frac{a_{4,17}}{a_{4,19}}
\end{align}
with $T_{i,k}$ the brightness temperature of horn $i=2,4$ at frequencies $k=17,19$\,GHz respectively. 
%Since the coefficients $a_{2/4,17/19}$ are constants, they can be extracted from the cross-correlation function sum. 
To obtain the experimental ratio, we used the values of the cross-correlation functions at lag 0 for each selected observation. These values correspond to the amplitude of the correlation peak. 
%Additionally, the cross-correlation functions were left unnormalized, as specified in Equation \ref{eq:ratio_exp}. 
The set of observations used here is the same as the one used in the CL analysis. 
%For each observation and azimuth, we hence obtained the ratio according to Equation \ref{eq:ratio_exp}. 
Figure~\ref{fig:median_ratio} shows the median ratio as a function of azimuth, using bins of $10^\circ$ in azimuth. The error bars are obtained from the standard error of the mean (SEM). The total median experimental ratio is $R=0.436\pm0.027$, which is fully consistent with the  predictions from {\sc am}.
This result confirms the underlying hypothesis in this analysis that all the correlated signal between horns 2 and 4, is due to atmospheric emission, and that the amplitude is better modelled with the {\sc am} code.
%

%--------------------------------------------------------------------
\section{Measurement of the structure function from QUIJOTE MFI wide survey observations}
\label{section6}

We use the QUIJOTE MFI wide survey data to characterise the atmospheric angular correlation via its structure function.
For a given angular scale $\theta$, the two-point correlation function can be computed as 
\begin{eqnarray}
    C(\theta) = \langle T(\theta_i) \cdot T(\theta_j) \rangle,
\end{eqnarray}
where the average is computed over all  pairs of elements $i$ and $j$ in the timeline with the same angular separation $\theta = \mid \overrightarrow{\rm\theta_j} - \overrightarrow{\rm\theta_i} \mid$, 
being  $\theta_i$ and $\theta_j$ their sky coordinates, and $T(\theta_{i})$ the corresponding antenna temperature for that sample. 
From here, the structure function $D(\theta)$, which quantifies the fluctuations of a signal over different scales, is defined as
\begin{align} 
\label{eq:struc_fct}
    D(\theta) &= \frac{1}{2}\langle (T(0) - T(\theta))^2 \rangle \\ \nonumber
    &= \langle T(0)^2 \rangle  - \langle T(0) T(\theta) \rangle \\  \nonumber
    &= C(0) - C(\theta),
\end{align}
with $T(0)$ and $T(\theta)$ being the antenna temperature of 
%samples with angular separation $0^\circ$ and $\theta$, respectively. 
%
two samples with angular separation $\theta$.
The Kolmogorov theory predicts the following scale dependence for the structure function, as derived in \cite{Morris2022}.

\begin{equation} \label{eq:struc_theo}
    C(0) - C(\theta) \propto \theta ^{5/3 \simeq 1.7}.
\end{equation}

%--------------------------------------------------------------------
%\subsection{Measurement of the structure function }

\begin{figure}
    \centering
    \includegraphics[width = 9cm]{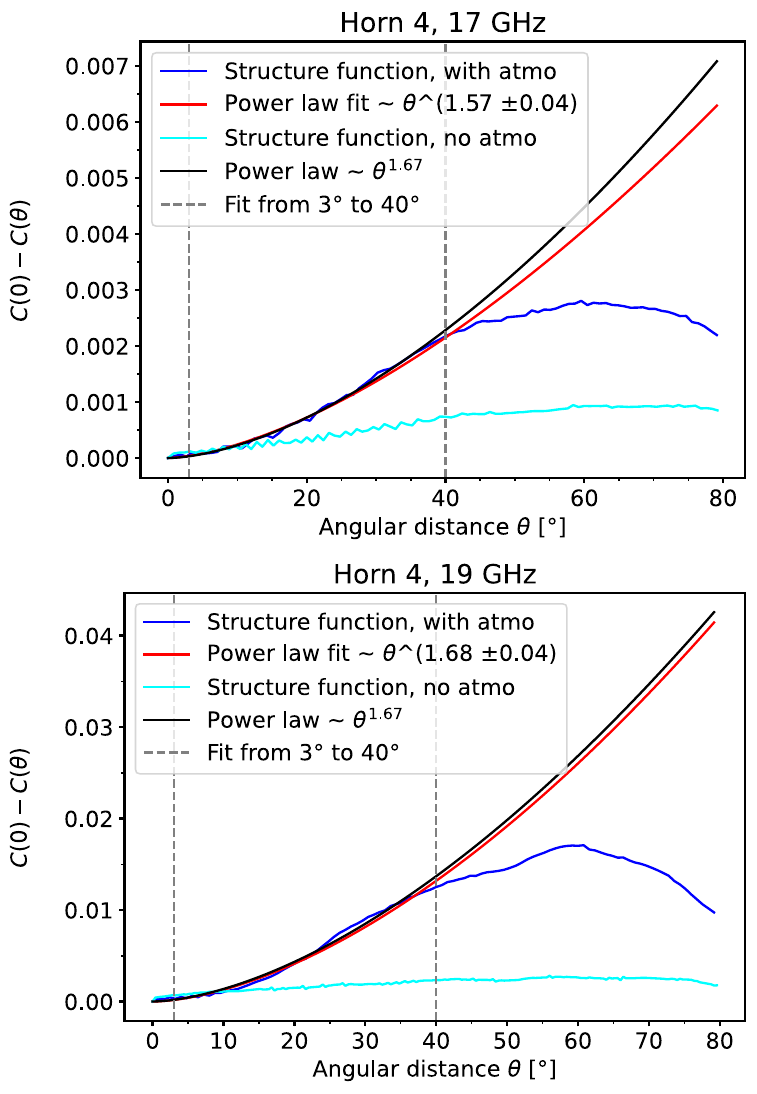}
    \caption{Structure function of the signal measured by horn 4 at 17\,GHz (top) and 19\,GHz (bottom) for the first 3.5 minutes of an observation of the MFI wide survey (the same observation as the one plotted in Figure \ref{fig:signals_comp}) calculated using equation \ref{eq:struc_fct}. The structure function calculated on the output signal is shown in deep blue, while the structure function calculated on data after atmospheric correction is shown in cyan. A power-law fit (red) was applied to the deep-blue curve over angular separations between 3° and 40°. The expected theoretical power law for the structure-function (equation \ref{eq:struc_theo}) is shown in black.}
    \label{fig:struc_fct}
\end{figure}

The structure function can be computed for the range of samples' angular separation of any single MFI observation. 
To guarantee that the timeline data is dominated by atmospheric emission and not instrumental noise, we pre-selected observations taken under high PWV values. For observations with low PWV, atmospheric noise blends with instrumental noise, significantly affecting the shape of the structure-function. 

Figure~\ref{fig:struc_fct} shows one example of the structure-function at 17\,GHz (top) and 19\,GHz (bottom) for the first 6.5\,min (10,000 samples) of an observation of the wide survey. The first 3 min of this observation were already shown in Figure \ref{fig:signals_comp}. 
We chose the first 6.5 minutes of the observation to calculate the structure function for several reasons. On the one hand, this ensured that we were observing the same atmospheric signal, which we have seen remains stable in time scales of 1--2 hours (see Section \ref{section:cross_corr}). On the other hand, this way we avoid heavy computations, as computation time increases drastically with the number of samples considered. For this observation, the telescope scanned the sky in a circular motion at an elevation of $50^\circ$; the maximal sample angular separation was hence $83^\circ$ \footnote{Note that we have quoted 50$^\circ$ as the elevation of the centre of the focal plane. The exact elevation of horn 4, which is located a bit below in the focal plane, is 48.2$^\circ$.}. The median PWV was $14.1$\,mm, which is high, and hence ensures a strong atmospheric signal. On Figure \ref{fig:struc_fct}, the deep blue curve shows the structure function of the output signal before atmospheric correction, i.e. with the atmospheric signal (grey curve on Figure \ref{fig:signals_comp}). The cyan curve shows the structure function of the signal after atmospheric correction using PCA (green curve on Figure \ref{fig:signals_comp}). 

The Kolmogorov behaviour is observed for the deep blue curve for angular separations below 40$^\circ$. Indeed, from the power law fit to the data with angular separations between 3$^\circ$ and 40$^\circ$ (solid red), we obtained power-law exponents of 1.57 at 17\ GHz and 1.68 at 19\ GHz, which are very close to the value of 1.67 predicted by the Kolmogorov theory. We interpret the flattening of the structure function beyond 40$^\circ$ as being due to the presence of the outer scale with this angular size. Indeed, this is comparable with the angular size of the outer scale found by \cite{Morris2024} in the Atacama desert using ACT data (see their Fig.~9). Using a scale height for the water vapour of 1\,km as found in Section~\ref{section3} (see Fig.~\ref{fig:med_rho}), this corresponds to an outer scale with a physical size of $\sim 700$\,m, also comparable to the findings of \cite{Morris2024}.

%However, the fact that the Kolmogorov behaviour is not retrieved at angular separations higher than $40^\circ$ could be explained by several facts. First, there are fewer sample pairs with higher angular separation, and this affects the statistics and, ultimately, the shape of the structure-function. Second, this structure-function was calculated using CTOD, for which the median intensity during each telescope scan was removed (grey curve in Figure \ref{fig:signals_comp}). This means that the median atmospheric signal is lost in each scan, strongly influencing sample pairs with high angular separation. Another effect explaining the departure from theory at high angular scales is the fact that the atmospheric conditions can vary significantly depending on the direction of observation. % I don't really know what to say more about this...

We also computed the structure function of the signals after subtraction of the atmospheric signal. This resulted in the complete deviation of Kolmogorov’s theory, as seen from the cyan curves in Figure \ref{fig:struc_fct}. This confirms that the atmospheric signal is dominating the instrumental noise for this dataset and that the atmospheric signal is responsible for the 1.67 power-law behaviour.  However, it is important to notice that this behaviour is only obtained for observations with very high PWV. When we computed the structure-function for observations taken during low PWV conditions (not shown in this paper for brevity), we obtained power-law exponents in the range of 0.8 to 0.9. This flattening is attributed to the contribution of instrumental noise, which follows a much flatter power-law slope compared to the atmosphere. 

%--------------------------------------------------------------------
\section{Characterization of the atmospheric power spectrum with MFI2 observations} \label{section:MFI2}

\begin{figure}
    \centering
    \includegraphics[width = 9cm]{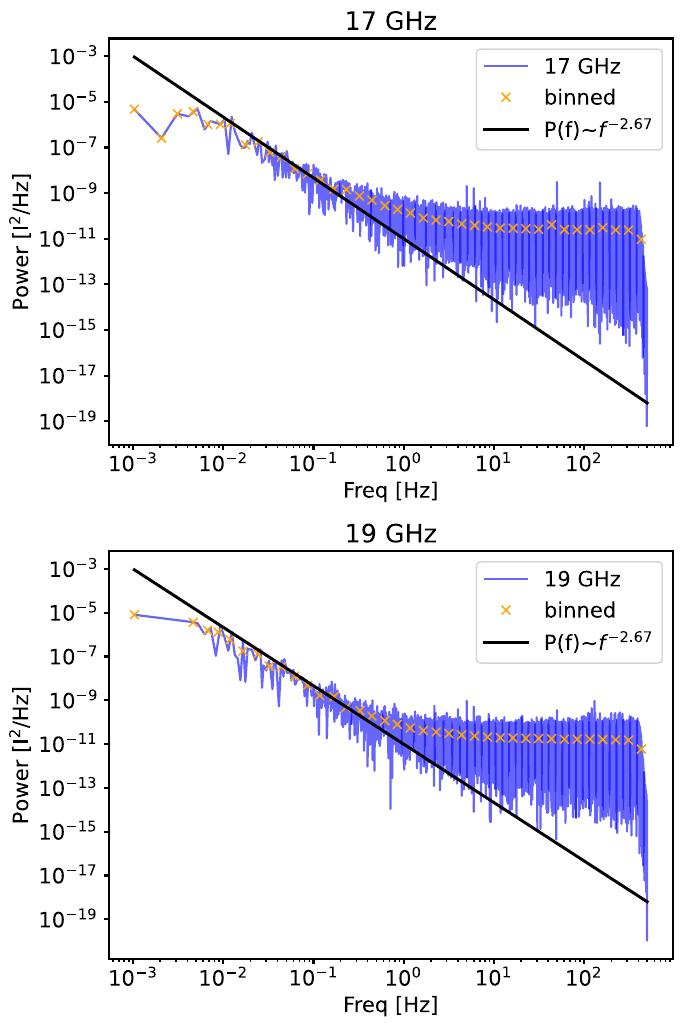}
    \caption{Cross power spectral density according to equation \ref{eq:CPSD} of the intensity signals of horns 2 and 4 of one dataset of MFI2 at 17\,GHz (top) and 19\,GHz (bottom). This dataset was taken on the 8th of July 2024 at 11:23 UTC+0 at a telescope elevation of $30^\circ$. The median wind speed during the observation was 4.53\,m/s and the median wind direction was 302° (North-West). The median PWV was 1.73\,mm. The CPSDs are plotted in blue, and the binned (40 bins) CPSDs are overplotted with orange crosses. The Kolmogorov spectrum is depicted in black.}
    \label{fig:MFI2_CPSD}
\end{figure}
% ATMOS30-240708-1123-000_combined.txt
% PWV=1.73mm
% Median wind direction: 302.03°
% Median wind speed: 4.53 m/s
% We see evidence of outer scale at 0.01\,Hz at both frequencies.

Next, we used MFI2 observations to characterise the atmospheric power spectrum and compare it with the theoretical predictions.
\citet{Tatarski1961} showed that the Kolmogorov theory \citep{kolmogorov1941} can be used to describe the atmospheric turbulence since water vapour evolves according to the velocity field in turbulent velocities.
In the inertial sub-range, the turbulence should follow a scale-invariant spectrum in 3D:
\begin{equation}
    \Phi(\kappa) \propto \kappa^{-11/3}, \quad 1/L_0 < \kappa < 1/l_0.
\end{equation}
where $\Phi(\kappa)$ is the energy density at that scale, $L_0$ is the size of the outer scale (the size of the largest turbulence eddies), and $l_0$ is the inner scale (smallest eddies). In 2D, this spectrum can be approximated by \citep{Church1995}
\begin{equation}
    \Phi(\kappa) \propto \kappa^{-8/3}, \quad 1/L_0 < \kappa < 1/l_0.
\end{equation}

As explained in section \ref{section:filtering}, the scanning strategy used in the MFI wide survey yields artificial power at the telescope’s rotation frequency and its harmonics if we directly use the timelines from the MFI wide survey. This complicates the interpretation of the power spectrum and leads us to conduct specific observations with the telescope at a fixed position.

We use the new MFI2 instrument for these observations. When the telescope is stationary, the atmospheric structure moves in front of the telescope due to the wind. This movement allows the instrument to probe the spatial structure and the turbulence of the atmosphere as it naturally flows in front of the horn's beams.

An important effect considered in this analysis is the instrumental $1/f$ noise present in each horn signal. This noise has a power spectrum that has a similar shape to that of the atmospheric signal, but with a different (much flatter) slope. As a result, when the instrumental noise is not corrected for, it flattens the overall slope of the atmospheric power spectrum. To eliminate this noise contribution, we computed the Cross Power Spectral Density (CPSD) between signals from different horns observing at the same frequency bands. Since each horn has its own independent amplifier chain, their instrumental noise contributions are uncorrelated and therefore do not appear in the CPSD, allowing us to isolate the atmospheric signal more effectively. The CPSD of the signals were computed according to
\begin{equation} \label{eq:CPSD}
    P_{h_2 h_4} = \frac{\mathcal{F}_{h_2} \cdot (\overline{\rm \mathcal{F}_{h_4}})}{f_s \cdot N}, 
\end{equation}

\noindent
with $\mathcal{F}_{h_2/h_4}$ the Fourier transform of the time domain signals of horns 2/4, $\overline{\rm \mathcal{F}}$ the complex conjugate, $f_s$ the signal sampling frequency and $N$ the number of samples. However, the correlation between the horns' signals will not be perfect since the horns are located 36\,cm apart on the focal plane. This results in the fact that they observe regions of the sky that are separated by approximately 5°. This effect will lead to partial decorrelation of the atmospheric signal at small scales. 

We computed the CPSD of horns 2 and 4 signals for the 19 MFI2 observations (see Table \ref{tab:list_MFI2} for the details of each observation), using equation \ref{eq:CPSD}. The CPSD for one observation is shown in Figure \ref{fig:MFI2_CPSD} at 17 GHz (top) and 19\, GHz (bottom) in a log-log plot. The slopes of the CPSD are consistent with Kolmogorov at both frequencies. Moreover, the flattening of the spectrum at low frequency also indicates the limit of the inertial sub-range regime and the outer scale $L_0$ of the turbulence at about $0.01$Hz. Given a median wind speed of $v=4.5$\,m/s during this observation, we can estimate $L_0$ of about $v/\kappa_0 = 450$\,m. This result is consistent with the size of the outer scale derived in \citet{Errard2015} and \citet{Morris2024} for the conditions of the Atacama desert. The flattening of the spectrum at high frequencies (above $\sim 0.1$\,Hz) is due to instrument white noise. Together with the presence of the outer scale at low frequencies, this limits the frequencies where the Kolmogorov spectrum is measured to a range of about $\sim 10$\,Hz. In this sense, \cite{Morris2022} benefited from the much larger number of detectors of ACT and by summing the signals measured by hundreds of them managed to lower the white-noise floor and thence measure the Kolmogorov slope over a wider frequency range.

%The CPSD of an 8 hours MFI observation is also shown in the Appendix, Figure \ref{fig:CPSD_no_med_one_obs}. This is the same observation as in Figure \ref{fig:signals_comp} and the one used to compute the structure-function in Figure \ref{eq:struc_fct}. On this CPSD, we also see the flattening occurring at a slightly lower frequency. However, this is consistent with $L_0 \approx 400$m since the median wind speed that day was about 1.5\,m/s. This observation is also in agreement with the dependence on the Kolmogorov power law.

% , as well as the CPSD of the averaged intensity for each observation

\begin{figure}[t]
    \centering
    \includegraphics[width = 9cm]{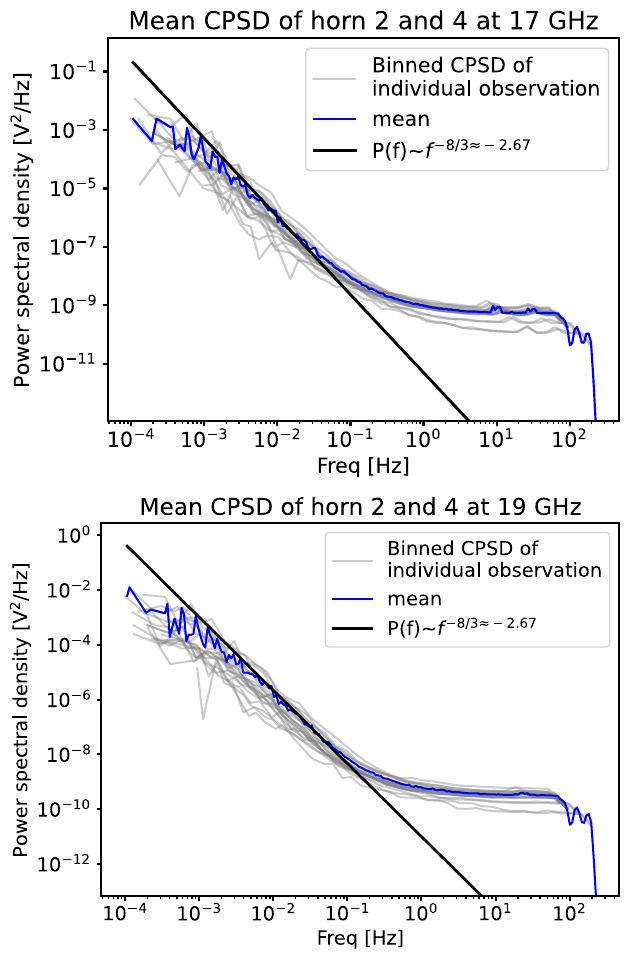}
    \caption{Averaged cross power spectral density (blue line) of the MFI2 intensity signal of horns 2 and 4 at 17\,GHz (top) and 19\,GHz (bottom) calculated using equation \ref{eq:CPSD}. The individual power spectral densities of the 19 observations are plotted as grey lines. The Kolmogorov spectrum is depicted in black.}
    \label{fig:averaged_CPSD}
\end{figure}

To obtain more statistically significant results, we computed the average CPSD of the 19 observations. Those results are presented in Figure \ref{fig:averaged_CPSD}. An effect to take into account when computing the average CPSD of the 19 spectra was the varying atmospheric conditions across different observations, namely, the different wind speeds. Those affect how spatial atmospheric fluctuations are translated into temporal fluctuations. If the wind is faster, the atmospheric fluctuations move quickly in front of the telescope, leading to higher-frequency variations in time. If the wind is slower, the variations occur more slowly. This is why we needed to normalise the spectra to a common reference. To do so, we applied the following scaling to the spectrum of each observation:

\begin{itemize}
    \item The frequency (x-axis) of each spectrum was divided by the median wind speed (values given in Table \ref{tab:list_MFI2}). 
    %This shifts the frequency axis so that observations with faster wind conditions are directly comparable to those with slower wind conditions.
    This way, we ensure that data affected by turbulence with a common spectrum in the spatial domain are combined coherently in the frequency domain.
    \item  The power (y-axis) of each spectrum was multiplied by the median wind speed squared. This is because the power of the fluctuations depends on the time the structures spend crossing the beam. Faster winds reduce this time, leading to lower apparent power.
\end{itemize}

\noindent
After rescaling each spectrum according to the median wind speed, we calculated the average spectrum by binning the power values and calculating the average power in each bin. The individual spectra of the 19 observations are plotted as light grey lines in Figure \ref{fig:averaged_CPSD}. The averaged CPSD is plotted in blue, and the Kolmogorov prediction in black. We see evidence that the spectrum follows Kolmogorov at both 17\,GHz and 19\,GHz, with again evidence of the impact of the outer scale below 0.01\,Hz.
%At 17\,GHz, the spectrum is slightly flatter than Kolmogorov. This might be due to the fainter atmospheric signal or other effects. 
However, it is difficult to assess with certainty, and further analysis is needed to confirm whether we are truly observing the Kolmogorov spectrum. For instance, the analysis could be repeated with a larger number of observations, or with another instrument such as the QUIJOTE Thirty-Forty Gigahertz Instrument (TFGI) \citep{TFGIreceivers_artal}, which has more horns. This would allow us to average over a larger number of closely spaced horns on the focal plane. This would help decrease the white noise as well as the $1/f$ noise thanks to the combination of different detectors with independent $1/f$ properties, in a similar way to the analysis applied to ACT data by \cite{Morris2022}.

%such as the wind direction

%--------------------------------------------------------------------
\section{Discussion and conclusion} \label{section8}

We have performed an analysis of the atmospheric signal at the Teide Observatory using various data sources.
%measured from QUIJOTE MFI and MFI2 in the frequency band 17\,GHz and 19\,GHz. 
First, we presented the average atmospheric conditions at the OT using data from radio-sounding launched from G\"uímar, as well as data from the GPS station and the STELLA observatory located at the OT. From those data sources, we obtained the median profiles of atmospheric temperature, water vapour density and atmospheric pressure, and the distribution of PWV, wind speed and wind direction during the QUIJOTE MFI wide survey.

The next part of the analysis focused on verifying the assumption that atmospheric conditions generally remain stable over a minimum time period of one hour during QUIJOTE MFI wide survey observations. This assumption was crucial for correcting atmospheric signals in the MFI wide survey. 
%To test it, we computed the cross-correlation function between signals from different horns to estimate the coherence length, that is, the width of the correlation peak, which provides insight into how long the signals remain correlated. 
We found that the median coherence length ranges between 1 and 2 hours, validating our initial assumption. These findings were complemented with the measurements of the correlation ratios between 17 and 19\,GHz. Our results are found to be in agreement with the theoretical prediction for those ratios based on {\sc am} model. 

Then, we used the observations from the QUIJOTE MFI wide survey to compute the angular structure function of the atmosphere. 
%and compared the results with Kolmogorov theory. 
Our results show good agreement with the Kolmogorov model prediction for those observations with strong atmospheric signals, i.e. high PWV content.
%, since in those cases, atmospheric fluctuations dominate over instrumental noise by far. 

Finally, we computed the cross-power spectral density between different horns using MFI2 observations, where the telescope remained in a fixed position. The spectral slope of the $1/f$ atmospheric emission is found to be in agreement with the Kolmogorov prediction of approximately $\sim 2.7$, although a slightly flatter index is found at 17\,GHz. We also find evidence for the presence of a turbulence outer scale on the order of 500 meters.

% FUTURE works. Maybe move to conclusions.
These results might have implications for future experiments such as GroundBIRD \citep{lee2020groundbird}, LSPE-STRIP \citep{addamo2021large}, the Tenerife Microwave Spectrometer (TMS) \citep{TMS}, and for extending QUIJOTE to higher frequency bands. 
More broadly, our analysis provides valuable insight for the wider CMB community next-generation of ground-based telescopes that require a deeper understanding of atmospheric contamination, such as the Simon Observatory (SO) \citep{SO} or the Cosmology Large Angular Scale Surveyor (CLASS) \citep{essinger2014class}. 
%

%--------------------------------------------------------------------
\begin{acknowledgements}
This work was supported by the Centre National de la Recherche Scientifique (CNRS), France, UMR8617 and Université Paris Saclay.
We thank the staff of the Teide Observatory for invaluable assistance in the commissioning and operation of QUIJOTE. The QUIJOTE experiment is being developed by the Instituto de Astrofisica de Canarias (IAC), the Instituto de Fisica de Cantabria (IFCA), and the Universities of Cantabria, Manchester and Cambridge. Partial financial support was provided by the Spanish Ministry of Science and Innovation (MCIN/AEI/10.13039/501100011033) under the projects AYA2007-68058-C03-01, AYA2007-68058-C03-02, AYA2010-21766-C03-01, AYA2010-21766-C03-02, AYA2014-60438-P, ESP2015-70646-C2-1-R, AYA2017-84185-P, ESP2017- 83921-C2-1-R, PGC2018-101814-B-I00, PID2019-110610RB-C21, PID2020-120514GB-I00, IACA13-3E-2336, IACA15-BE-3707, EQC2018-004918-P, PID2023-150398NB-I00 and PID2023-151567NB-I00, the Severo Ochoa Programs SEV-2015-0548 and CEX2019-000920-S, the Maria de Maeztu Program MDM-2017-0765, and by the Consolider-Ingenio project CSD2010-00064 (EPI: Exploring the Physics of Inflation). We acknowledge support from the ACIISI, Consejeria de Economia, Conocimiento y Empleo del Gobierno de Canarias and the European Regional Development Fund (ERDF) under grant with reference ProID2020010108, and Red de Investigación RED2022-134715-T funded by MCIN/AEI/10.13039/501100011033. This project has received funding from the European Union’s Horizon 2020 research and innovation program under grant agreement number 687312 (RADIOFOREGROUNDS), and the Horizon Europe research and innovation program under GA 101135036 (RadioForegroundsPlus).
We acknowledge the use of the \texttt{SciPy}, \texttt{numpy}~\citep{numpy}, and \texttt{matplotlib}~\citep{matplotlib} software packages. 
\end{acknowledgements}

\bibliographystyle{aa}
\bibliography{aanda}

%\newpage

% include the Appendix
\include{Appendix}


\end{document}

%% file: Appendix.tex
\appendix

\section{Observation parameters for MFI2} \label{ap_MFI1}

The basic parameters of the 19 QUIJOTE MFI2 observations used in Sect.~\ref{section:MFI2} are listed in Table~\ref{tab:list_MFI2}. These observations were employed to compute the atmospheric power spectral density.

\begin{table*}
\centering
\begin{tabular}{ccccccc}
\hline
Elevation & $T_\text{observed}$ & Date  & Time UTC  & PWV & Wind speed & Wind direction\\ 
(deg) & (min) & (dd/mn/yy) & (hh:mm) & (mm) & (m/s) &  (deg) \\ 
\hline \hline 
         30 & 21.08 & 02/07/2024 & 12:09 & 5.45 & 3.14 & 232.57 \\ 
         30 & 16.08 & 06/07/2024 & 11:31 & 4.22 & 5.07 & 248.37 \\
         30 & 16.08 & 07/07/2024 & 11:27 & 1.07 & 4.62 & 300.77 \\
         30 & 16.08 & 08/07/2024 & 11:23 & 1.73 & 4.53 & 302.03 \\
         30 & 16.08 & 09/07/2024 & 11:19 & 0.81 & 4.90 & 309.55 \\
         30 & 16.08 & 10/07/2024 & 11:15 & 0.03 & 5.99 & 57.08 \\ \hline
         60 & 18.08 & 02/07/2024 & 11:50 & 4.07 & 3.26 & 257.77 \\
         60 & 18.08 & 06/07/2024 & 11:48 & 4.22 & 5.07 & 226.85 \\
         60 & 18.08 & 07/07/2024 & 11:44 & 1.07 & 4.62 & 299.83 \\
         60 & 18.08 & 08/07/2024 & 11:40 & 1.73 & 4.53 & 300.36 \\
         60 & 18.08 & 09/07/2024 & 11:36 & 0.81 & 4.90 & 286.60 \\
         60 & 18.08 & 10/07/2024 & 11:32 & 0.03 & 5.99 & 62.09 \\
         60 & 20.08 & 11/07/2024 & 11:13 & 5.69 & 2.21 & 62.09 \\ \hline
         90 & 28.10 & 06/07/2024 & 12:07 & 6.14 & 5.26 & 235.44\\
         90 & 28.10 & 07/07/2024 & 12:03 & 2.61 & 4.68 & 305.45 \\
         90 & 28.10 & 08/07/2024 & 11:59 & 1.73 & 4.53 & 300.42 \\
         90 & 26.10 & 09/07/2024 & 11:55 & 0.81 & 4.90 & 301.24 \\
         90 & 28.10 & 10/07/2024 & 11:51 & 0.03 & 5.99 & 62.01 \\
         90 & 24.10 & 11/07/2024 & 11:34 & 5.69 & 2.21 & 142.70 \\
         \hline\hline 
    \end{tabular}
    \caption{List of observations taken with the QUIJOTE MFI2 instrument for this study. Column 1 indicates the elevation. Column 2 lists the duration of each observation in minutes, while columns 3 and 4 give the date and time of the observations. Column 5 shows the median PWV during each observation measured by the Izaña weather station (see section \ref{section:Izana} for details). Columns 6 and 7 show the median wind speed and median wind direction during the observation measured by STELLA weather station (see section \ref{section:Stella} for details).}
    \label{tab:list_MFI2}
\end{table*}

\section{MFI observation selection for the coherence length analysis} \label{section:data_sel}

\begin{table}
    \centering
    \begin{tabular}{cccc}
    \hline
    Azimuth & Obs.  & Obs. & Obs. \\
    ($^\circ$) &  removed   &  removed (\%) &  kept  \\ 
    \hline  \hline 
    0 & 161 & 30.84 & 361 \\
    10 & 155 & 29.69 & 367 \\
    20 & 183 & 35.06 & 339 \\
    30 & 211 & 40.42 & 311 \\
    40 & 184 & 35.25 & 338 \\
    50 & 233 & 44.64 & 289 \\
    60 & 196 & 37.55 & 326 \\
    70 & 161 & 30.84 & 361 \\
    80 & 159 & 30.46 & 363 \\
    90 & 157 & 30.08 & 365 \\
    100 & 140 & 26.82 & 382 \\
    110 & 190 & 36.40 & 332 \\
    120 & 249 & 47.70 & 273 \\
    130 & 218 & 41.76 & 304 \\
    140 & 186 & 35.63 & 336 \\
    150 & 184 & 35.25 & 338 \\
    160 & 255 & 48.85 & 267 \\
    170 & 315 & 60.34 & 207 \\
    180 & 251 & 48.08 & 271 \\
    190 & 287 & 54.98 & 235 \\
    200 & 193 & 36.97 & 329 \\
    210 & 291 & 55.75 & 231 \\
    220 & 218 & 41.76 & 304 \\
    230 & 255 & 48.85 & 267 \\
    240 & 233 & 44.64 & 289 \\
    250 & 218 & 41.76 & 304 \\
    260 & 246 & 47.13 & 276 \\
    270 & 157 & 30.08 & 365 \\
    280 & 161 & 30.84 & 361 \\
    290 & 154 & 29.50 & 368 \\
    300 & 160 & 30.65 & 362 \\
    310 & 160 & 30.65 & 362 \\
    320 & 140 & 26.82 & 382 \\
    330 & 137 & 26.25 & 385 \\
    340 & 168 & 32.18 & 354 \\
    350 & 170 & 32.57 & 352 \\ \hline
    Median: & 185 & 35.44  &  337 \\ \hline \hline 
    \end{tabular}
    \caption{Number of discarded QUIJOTE MFI wide-survey observations at each azimuth, out of a total of 522. Column 1 lists the azimuth values considered. Column 2 gives the number of observations removed after applying conditions 1, 2, and 3, with the corresponding percentages shown in Column 3. Column 4 indicates the number of observations retained for the final analysis. The last row of the table reports the median value for each column.}
    \label{tab:nb_datasets}
\end{table}

%\jarm{CHECK}

For the coherence length analysis, we selected a subset of QUIJOTE MFI wide survey observations from the 1223 used to build the public maps \citep{mfiwidesurvey}. We first discarded observations with an excessive number of flagged samples, and subsequently excluded those for which the automated Gaussian fitting procedure did not yield satisfactory results.

\subsection{Observations removed due to flagging}
\begin{figure}
    \centering
    \includegraphics[width = 9cm]{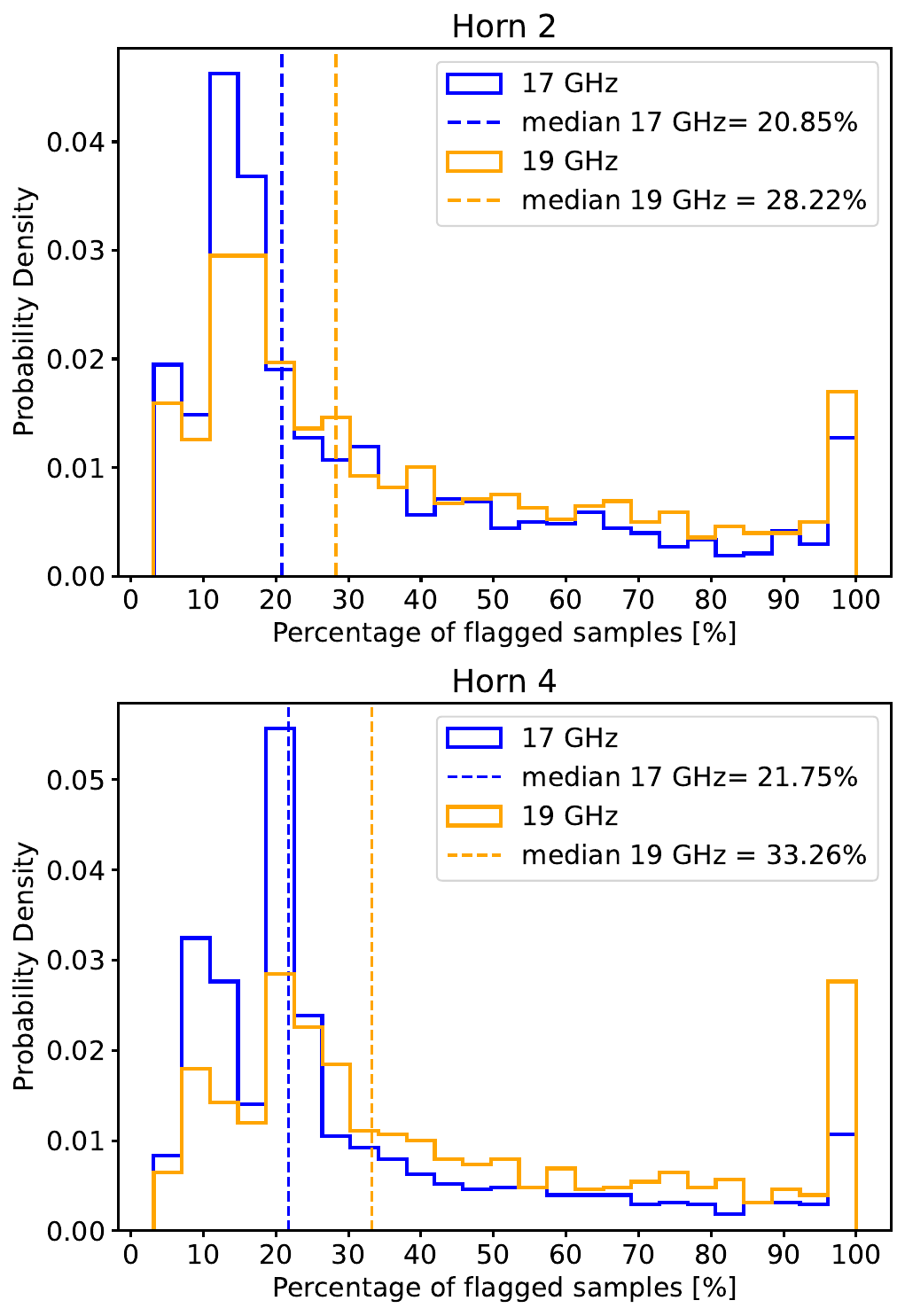}
    \caption{Distribution of the percentage of flagged samples in the observations of the QUIJOTE MFI survey for horn 2 (top) and horn 4 (bottom) at 17\,GHz (blue) and 19\,GHz (orange).}
    \label{fig:perc_zeroes}
\end{figure}

We analysed the fraction of flagged samples in each QUIJOTE MFI observation and selected those observations with less than 30\,\% of flagged data. Figure~\ref{fig:perc_zeroes} shows the distribution of the percentage of flagged samples in the CTOD of horn 2 (top) and horn 4 (bottom) at 17\,GHz (blue) and 19\,GHz (orange). For horn 2, the median fraction of flagged samples in the wide survey observations is 21\,\% at 17\,GHz and 28\,\% at 19\,GHz, while for horn 4 it is 22\,\% and 33\,\%, respectively. %The higher fraction at 19\,GHz is mainly due to increased satellite emission in this frequency band. 
After applying this selection, 522 out of 1233 observations were retained.

\subsection{Observations removed due to poor Gaussian fitting process}

We also discarded observations for which the Gaussian fitting process did not yield satisfactory results. The following conditions were considered:
 
\begin{enumerate}     
     \item The fit reliability condition: the goodness of fit, measured by the coefficient of determination $R^2$, must be greater than 0.8:     
     \begin{equation}\label{eq:goodness_fit}
         R^2 = 1 - \frac{SS_{res}}{SS_{tot}} > 0.8,
     \end{equation}

     \noindent
     with $SS_{res}$ the sum of squares of residuals is defined as:
     \begin{equation}
         SS_{res} = \sum(y_i - f_i)^2,
     \end{equation}
     
     \noindent
     with $f_i$ the $i$th fitted values, $y_i$ the $i$th data point, and $SS_{tot}$ the total sum of squares defined as:     
     \begin{equation}
         SS_{tot} = \sum(y_i - \overline{y})^2,
     \end{equation}
     
     \noindent 
     with $\overline{y}$ the mean of the data. Furthermore, the root-mean-square error (RMSE) of the fit was calculated to ensure reliable results.     
     %\begin{equation} \label{eq:RMSE}
     %    \text{RMSE} = \sqrt{<\text{residual}^2>}.
     %\end{equation}
     
     \item The fit parameters condition: the amplitude of the Gaussian fit $a_{fit}$ must be positive to avoid fitting anti-correlation. Furthermore, the standard deviation of the Gaussian  $\sigma_{fit}$ must be larger than $0.8 $\,s to prevent fitting correlation peaks caused by the telescope's scanning rotation.
     
    %\begin{eqnarray}
    %    &&a_{fit} > 0, \\
    %    &&\sigma_{fit} > 0.8\text{s}.
    %\end{eqnarray}

     \item Condition on the number of samples: 
     the observation must contain more than 25,000 samples (each of 40\,ms). Files with fewer samples were discarded, as they correspond to observation times shorter than 16 minutes.
 \end{enumerate}
 
Table~\ref{tab:nb_datasets} gives the number of discarded observations for each azimuth out of the 522 initially retained after flagging criteria, based on the goodness-of-fit and minimum number of samples criteria. It also gives the corresponding percentage of removed observations and the final number of observations used in this analysis at each azimuth.

%Figure \ref{fig:CL_bad_fit} shows the cross-correlation function for four observations that were removed according to the goodness-of-fit criteria. It also explained why the criteria for those observations were not met.

\begin{figure}
\centering
\includegraphics[width = 9cm]{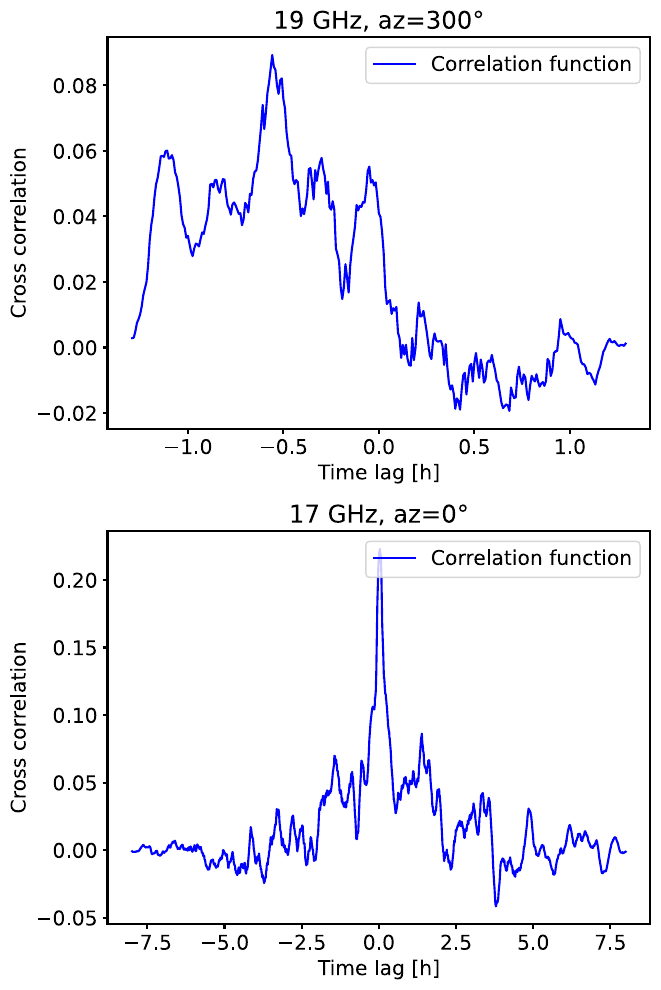}
    \caption{Examples of discarded observations. Top panel: normalised cross-correlation function between the signals from horns 2 and 4 at 19\,GHz for an observation at an azimuth of  $300^\circ$. Bottom panel: normalised cross-correlation function between the signals from horns 2 and 4 at 17\,GHz for another observation excluded from the analysis at an azimuth of $0^\circ$. See the text for details.}
\label{fig:CL_bad_fit}
\end{figure}

Figure~\ref{fig:CL_bad_fit} shows examples of normalised cross-correlation functions of observations that were excluded from the analysis. So observations that were removed because of condition 1, 2 or 3. The top panel shows the cross correlation at 19\,GHz of an observation taken on the 25th of June 2013 at 05:55 UCT+0, at azimuth $300^\circ$. This observation was excluded because the goodness of the fit at 19\,GHz was not high enough ($R^2<0.8$ according to equation \ref{eq:goodness_fit}). On this cross-correlation function, we do not see a clear correlation peak at 0\,h time lag, and the correlation is overall small, of the order 
of 8\,\%, indicating a weak atmospheric signal. Indeed, the PWV was 1.1\,mm that day, which would produce a weak atmospheric signal, and hence a small correlation between the signals of horns 2 and 4. The bottom panel shows the cross correlation at 17\,GHz for an observation taken on the 30th of August of 2016 at 14:45 UCT+0 at azimuth $0^\circ$. The PWV was 14.1\,mm. It was excluded for the same reason as the top panel observation, i.e. insufficient goodness of fit for one of the observations, either at 17 or 19\,GHz. 
In this case, the issue occurred at 17\,GHz. Although the correlation function shows a clear correlation peak (around 22\,\% correlation between the two horns’ signals after smoothing), the peak does not exhibit a Gaussian profile, making it difficult for the automated fitting process to converge properly. Consequently, the fit quality was not sufficient, which explains why this observation was discarded at this azimuth.

